\documentclass[preprint]{elsarticle}
\usepackage[latin9]{inputenc}
\usepackage{array}
\usepackage{amsmath}
\usepackage{amssymb}
\usepackage{graphicx}
\PassOptionsToPackage{normalem}{ulem}
\usepackage{ulem}

\makeatletter

\providecommand{\tabularnewline}{\\}

\usepackage{amsfonts}
\usepackage{multirow}
\usepackage{url}
\providecommand{\U}[1]{\protect \rule{.1in}{.1in}}

\makeatother

\begin{document}
\begin{frontmatter}

\title{Design Automation and Optimization Methodology for Electric Multicopter
UAVs}

\author{Xunhua Dai\textsuperscript{a}}

\author{Quan Quan\textsuperscript{a}}

\author{Kai-Yuan Cai\textsuperscript{a}}

\address{\textsuperscript{a}School of Automation Science and Electrical Engineering,
Beihang University, Beijing, 100191, China}
\begin{abstract}
The traditional multicopter design method usually requires a long
iterative process to find the optimal design based on given performance
requirements. The method is uneconomical and inefficient. In this
paper, a practical method is proposed to automatically calculate the
optimal multicopter design according to the given design requirements
including flight time, altitude, payload capacity, and maneuverability.
The proposed method contains two algorithms: an offline algorithm
and an online algorithm. The offline algorithm finds the optimal components
(propeller and electronic speed controller) for each motor to establish
its component combination, and subsequently, these component combinations
and their key performance parameters are stored in a combination database.
The online algorithm obtains the multicopter design results that satisfy
the given requirements by searching through the component combinations
in the database and calculating the optimal parameters for the battery
and airframe. Subsequently, these requirement-satisfied multicopter
design results are obtained and sorted according to an objective function
that contains evaluation indexes including size, weight, performance,
and practicability. The proposed method has the advantages of high
precision and quick calculating speed because parameter calibrations
and time-consuming calculations are completed offline. Experiments
are performed to validate the effectiveness and practicality of the
proposed method. Comparisons with the brutal search method and other
design methods demonstrate the efficiency of the proposed method.
\end{abstract}



\end{frontmatter}

\section{Introduction}

Unmanned aerial vehicles (UAVs) are widely applied to increasingly
new fields and environments \citep{Dai2018}, especially electric
multicopters. However, designing a multicopter for specific commercial
use is not an easy task because the performance and efficiency of
a multicopter are highly coupled with the actual environment (e.g.,
altitude and temperature) and the flight conditions (e.g., payload
weight, battery condition, and reserve thrust). For example, although
a consumer multicopter (e.g., the DJI$^{\circledR}$ Phantom) can
fly flexibly and efficiently in its normal condition, its maneuverability
and efficiency are significantly reduced in high-latitude environments
(low air density and temperature) or high-payload flight conditions. 

In these cases, the multicopter may be unable to take off normally,
so the multicopter must be redesigned by changing the propulsion system
(usually the propeller) or reducing the payload weight. For commercial
multicopters (e.g., logistical and plant protection multicopters),
the actual environment and flight conditions are more complicated,
thus requiring more design schemes to ensure normal operations. In
practice, even for experienced designers, multiple trial-and-error
experiments are required to design, verify, and update the multicopter
design according to the given requirements. This is a costly and time-consuming
process. Consequently, we want to present a practical method to automatically
calculate the optimal multicopter design for different task requirements.
This will improve the multicopter design efficiency and broaden its
application scenarios. 

\begin{figure}[tbh]
\centering \includegraphics[width=0.65\textwidth]{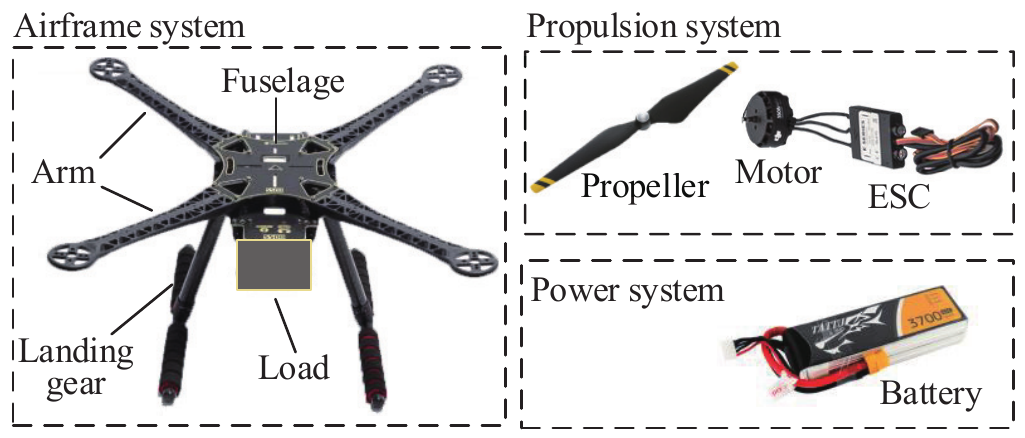}\caption{Typical compositions of multicopter body systems \citep{quan2017introduction}.}
\label{Fig01} 
\end{figure}

A multicopter system can be divided into a body system and a flight
control system, and a multicopter design optimization problem usually
refers to the design of the body system. As shown in Fig.\,\ref{Fig01},
the body system can be further divided into three subsystems: the
propulsion system, which consists of a propeller, motor, and electronic
speed controller (ESC); the airframe system, which consists of several
arms, a fuselage, payload, and landing gear; and the power system,
which consists of a battery pack. 

There are thousands of available products on the market for the above
components to assemble a multicopter, for which it is difficult to
use traditional trial-and-error methods to find the optimal design
and component combination that satisfies all given requirements. In
addition, design constraints such as the safety and compatibility
of multicopters are too complex to describe with mathematical expressions.
A good multicopter design should be evaluated with considering more
factors such as performance (flight time, payload capability, and
maneuverability), cost, and size. However, it is difficult to find
a uniform standard to evaluate different multicopter designs. The
above difficulties make the design automation and optimization problem
difficult, and have led to only a few research studies in this field. 

The design automation and optimization methodology can not only improve
the design efficiency of the body system but also significantly shorten
the development period of the entire multicopter system. The model-based
development method \citep{ke2018design,santamaria2012model} is an
efficient way to accelerate the development of control systems, but
it requires a precise model of the multicopter, which depends on the
body system design result. According to \citep{oktay2016simultaneous},
designing the body system and control system of a UAV simultaneously
is an efficient way in the future. Subsequently, the obtained multicopter
system can be agile to respond to future changes in requirements.

Many optimization methods have been proposed for fixed-wing aircraft
\citep{Riboldi2018,Zhang2019} and helicopters \citep{Yeo2019}, but
the design optimization research for multicopters is scarce. Most
studies focused on optimizing the designs of specific components (e.g.,
motors, propellers, airfoil shapes, and fuselage aerodynamics) to
achieve longer flight endurance \citep{VolkanPehlivanoglu2019,Zhang2019,Vu2019}.
For example, a propeller efficiency analysis and optimization methods
are studied in \citep{Introduction2011,Deters2014}, and a motor efficiency
analysis and design methods are studied in \citep{burt2008electric}.
However, according to \citep{quan2017introduction}, the efficiency
of a multicopter is not only determined by the individual efficiency
of a motor or a propeller, but also by many other factors such as
the match between motors and propellers, the multicopter weight, and
the flight conditions. For example, an unmatched propeller may make
the motor work under a very inefficient state, which reduces the efficiency
of the entire multicopter. In another example, an efficient multicopter
may become very inefficient when the altitude (air density) or the
payload weight are changed. Therefore, the multicopter should be treated
as a whole to evaluate its performance and optimize its design.

In our earlier work \citep{Shi2017}, a practical method was proposed
to estimate the flight performance of a multicopter with a given configuration,
component selection, and flight conditions. This method allows designers
to verify and update the multicopter design according to the estimated
performance. The multicopter design optimization problem was essentially
a reverse process of the performance estimation problem in \citep{Shi2017}.
In our latest work \citep{dai2018apractical,dai2018EFF}, as the preliminary
work of the design optimization problem, an analytic method was proposed
to estimate the optimal parameters of propulsion systems in some simple
design tasks. Based on our previous studies, this paper focuses on
optimizing the multicopter as a whole by considering more practical
factors. The study in this paper is more difficult and meaningful
than our previous studies because the solution to the reverse process
is not analytical and must correspond to real products and flight
constraints.

A feasible way to solve the multicopter design optimization problem
is to use numerical searching methods to traverse all possible design
combinations and find the optimal one according to the objective functions
and constraints. For example, in \citep{Magnussen2014,Magnussen2015},
the multicopter optimization problem was described as a mixed-integer
linear program and solved with the CPLEX optimizer. In \citep{arellano2016multirotor},
the design optimization problem was solved based on discrete-integer
and continuous variables with the objectives of efficiency and flight
time. In \citep{tian2016mechatronic}, based on knowledge-based engineering
techniques, a design method was proposed to find a multicopter design
with the minimum cost and maximum payload. 

However, there exist some common problems with these methods. First,
the calculation speed is slow owing to the large computational cost
of traversing most of the possible combinations, especially when the
product database is large. For example, it took 5--25 s to find the
optimal design using a small-scale database (five motors and five
propellers) for the methods in \citep{Magnussen2014,Magnussen2015}.
However, in practice, a large product database is essential to cover
common multicopters with different weights and sizes, which means
that hours or days are required for these methods to obtain a reasonable
result. Second, the performance of the obtained designs is calculated
by theoretical estimation whose precision highly depends on the modeling
accuracy and parameter precision. Third, the user requirements and
optimization constraints considered in these papers are not comprehensive,
so it is difficult to use the obtained results to assemble a real
multicopter.

For the disadvantages of the existing methods, in this paper, a practical
method is proposed to solve the multicopter design optimization problem
with high computation speed and precision. The method contains two
algorithms: an offline algorithm and an online algorithm. The offline
algorithm finds the optimal components (propeller and ESC) for each
motor to obtain motor-ESC-propeller propulsion combinations. Subsequently,
all obtained propulsion combinations and their key performance parameters
are stored in a combination database. The online algorithm obtains
the multicopter design results that satisfy the given requirements
by searching through component combinations in the database and calculating
the optimal parameters for the battery and airframe. Subsequently,
these requirement-satisfied multicopter design results are evaluated
and sorted according to an objective function that contains evaluation
indexes including size, weight, performance, and practicability. Finally,
the optimal multicopter design is obtained according to the sorted
list.

Compared with other methods, the advantages of the proposed method
are as follows. (1) The calculation speed of the proposed method is
much faster because most of the time-consuming calculations are completed
offline. (2) The precision is higher because the parameters of the
combination database can be calibrated by experimental data. (3) The
obtained results are more practical because more design requirements
and constraints are considered in our method.

The proposed method is beneficial for improving the development speed
and reducing the cost of the design and verification of multicopters.
Experiments and comparisons demonstrate the effectiveness and practicality
of the proposed method. In addition, the method was published as an
online toolbox to provide users with a multicopter optimization design
service, and feedback indicates that the obtained results are accurate
and practical for multicopter designs.

The rest of the paper is organized as follows. \textsl{Section \ref{Sec-2}}
provides a comprehensive description of the multicopter design optimization
problem. In \textsl{Section \ref{SEC3}}, the offline algorithm is
presented to obtain a database with the optimal propulsion combinations.
In \textsl{Section \ref{sec:4}}, the online algorithm is presented
to obtain the optimal multicopter design. Experiments and comparisons
are discussed in \textsl{Section \ref{Sec-5}}. \textsl{Section \ref{Sec-6}}
presents the conclusions and future work.

\section{Problem Formulation}

\label{Sec-2}

This section presents the multicopter design optimization problem
from five aspects: inputs, outputs, constraints, objectives, and solving
method.

\subsection{Design Requirements}

The design requirements define the desired performance of a multicopter
under the given flight conditions. Based on a market survey of multicopters,
six important factors are considered in this paper: 

(1) \textsl{Hovering Time}. The hovering time $\hat{t}_{\text{fly}}$
(unit: min) is defined as the battery discharge time in the hovering
mode, in which the multicopter stays fixed in the air, and relatively
static to the ground.

(2) \textit{Payload Capability}. The payload weight $\hat{m}_{\text{load}}$
(unit: kg) describes the capacity to carry an instrument for certain
missions.

(3) \textsl{Maneuverability}. The maneuverability usually refers to
the acceleration ability, which is described by the thrust ratio $\hat{\gamma}\in\left(0,1\right)$
as
\begin{equation}
\hat{\gamma}\triangleq\frac{T_{\text{hover}}}{T^{*}}\label{eq:gama}
\end{equation}
where $T_{\text{hover}}$ (unit: N) is the hovering thrust of a single
propeller in the hovering mode, and $T^{*}$ (unit: N) is the full-throttle
thrust of a single propeller in the maximum control mode. According
to the force balance principle, the total propeller thrust is equal
to the multicopter weight in the hovering mode, which gives
\begin{equation}
n_{\text{p}}\cdot T_{\text{hover}}=m_{\text{copter}}\cdot g\label{eq:acc1}
\end{equation}
where $m_{\text{copter}}$ (unit: kg) is the multicopter mass, $g=9.8\text{m/s}^{2}$
is the acceleration of gravity and $n_{\text{p}}$ is the number of
propellers. Meanwhile, the vertical acceleration range of a multicopter
is $\left[-g,a_{\text{vMax}}\right]$, where $-g$ denotes the acceleration
in the free fall motion when all propellers stop rotating, and $a_{\text{vMax}}$
(unit: m/$\text{s}^{2}$) denotes the maximum vertical acceleration
when all propellers generate the full-throttle thrust $T^{*}$. According
to Eqs.\,(\ref{eq:gama}) and (\ref{eq:acc1}), $a_{\text{vMax}}$
is obtained as
\begin{equation}
a_{\text{vMax}}=\frac{n_{\text{p}}T^{*}-m_{\text{copter}}g}{m_{\text{copter}}}=\left(\frac{1}{\hat{\gamma}}-1\right)g.\label{eq:acc2}
\end{equation}
It can be observed from Eq.\,(\ref{eq:acc2}) that the acceleration
range (maneuverability) is directly determined by the thrust ratio
$\hat{\gamma}$. In practice, multicopters with different usages have
different maneuverability preferences. For example, racing multicopters
expect a smaller thrust ratio $\hat{\gamma}$ for higher maneuverability,
while load-carrying multicopters expect a larger $\hat{\gamma}$ to
allocate more weight for payload. In particular, when $\hat{\gamma}=0.5$,
the acceleration range obtained from Eq.\,(\ref{eq:acc2}) is $\left[-g,g\right]$,
which is widely adopted by designers because it has the most balanced
control range. Moreover, according to \citep{Shi2017,orsag2012influence},
the maximum flight speed of a multicopter can also be obtained by
the thrust ratio $\hat{\gamma}$ with the given aerodynamic coefficients.

\begin{figure}[tbh]
\centering \includegraphics[width=0.55\textwidth]{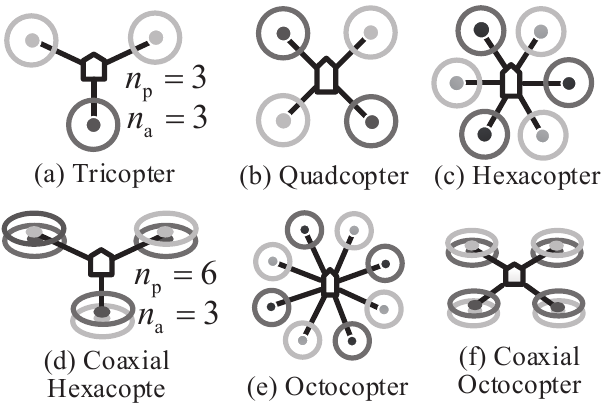}\caption{Basic airframe layouts for multicopters.}
\label{Fig02}
\end{figure}

(4) \textit{Airframe Layout}. According to \citep{quan2017introduction},
multicopter airframe layouts can usually be divided into common layouts
and coaxial layouts. Fig.\,\ref{Fig02} shows several basic airframe
layouts for multicopters, where Figs.\,\ref{Fig02}(a)(b)(c) are
of the common form and Figs.\,\ref{Fig02}(a)(b)(c) are of the coaxial
form. For common multicopters, the propeller number $n_{\text{p}}$
is equal to the arm number, and the propellers are assumed not to
interfere with each other. For coaxial multicopters, there are two
propellers on the same arm, and the propeller thrust efficiency may
decrease by about 20\% owing to the interaction between propellers
\citep{quan2017introduction}. This paper studies the design for common-layout
multicopters, and the proposed method can be easily applied to the
coaxial layout or other special layouts.

(5) \textit{Air Density (Altitude)}. The air density $\hat{\rho}$
(unit: kg/$\text{m}^{3}$) has a great influence on the performance
of a multicopter. For example, most multicopters cannot even take
off in high-altitude areas because the thrust becomes much smaller
than that in sea-level areas owing to the reduction of air density.
According to the international standard atmosphere (ISA) statistical
model \citep{cavcar2000international}, the air density can be estimated
by the altitude $h_{\text{t}}$ (unit: m) as
\begin{equation}
\begin{array}{cl}
\hat{\rho} & \approx\frac{273}{(273+T_{\text{t}})}(1-0.0065\frac{h_{\text{t}}}{273+T_{\text{t}}})^{5.2561}\rho_{\text{0}}\\
T_{\text{t}} & \approx T_{\text{0}}-\left(6\cdot h_{\text{t}}/1000\right)
\end{array}\label{eq:rho}
\end{equation}
where $\rho_{\text{0}}=1.293$kg/m$^{3}$ is the standard air density,
$T_{\text{t}}$ (unit: $^{\circ}$C) is the air temperature, and $T_{\text{0}}=25{}^{\circ}\text{C}$
is the average ground temperature.

(6) \textsl{Battery Type (Power Density)}. The power density $\rho_{\text{b}}$
(unit: W$\cdot$h/kg) determines the battery weight under the same
capacity. Although a battery with a higher power density is always
better in multicopter design, in practice, the selection of the battery
also depends on other factors such as safety, cost, and performance.
For a certain type of battery product, the power density is approximately
equal to a constant value. For example, for the most commonly used
Li-Po batteries, the power density is $\rho_{\text{b}}$$\approx240$W$\cdot$h/kg.

For simplicity, the above design requirements are marked with an input
parameter set $\mathbf{\Theta}_{\text{in}}$ as
\begin{equation}
\mathbf{\Theta}_{\text{in}}\triangleq\left\{ \hat{t}_{\text{fly}},\hat{m}_{\text{load}},\hat{\gamma},n_{\text{p}},\hat{\rho},\rho_{\text{b}}\right\} .\label{eq:PhiIn}
\end{equation}

\subsection{Design Outputs}

The design outputs contain all necessary information to assemble a
multicopter. 

(1) \textit{Product Selection of Propulsion System}. It is expected
that the proposed method can comprehensively determine the product
selection (motor, ESC, and propeller) of the propulsion system. Therefore,
a certain number of candidate products is necessary to be prepared
as the searching scope. For simplicity, the name and parameters (e.g.,
size, weight, and current limit) of a motor product are marked with
a parameter set $\Theta_{\text{m}}$. The motor database, which includes
all available motor products, is represented by a symbol $\mathbf{\Phi}_{\text{m}}$,
where $\Theta_{\text{m}}\in\mathbf{\Phi}_{\text{m}}$. Similar definitions
are applied to ESCs $\Theta_{\text{e}}\in\mathbf{\Phi}_{\text{e}}$
and propellers $\Theta_{\text{p}}\in\mathbf{\Phi}_{\text{p}}$. Next,
a propulsion system combination can be represented by $\left\{ \Theta_{\text{m}},\Theta_{\text{e}},\Theta_{\text{p}}\right\} $.

(2) \textit{Battery Parameters}. The battery pack is a highly customizable
component. Users can connect many small battery cells in serial and
parallel to obtain the desired voltage and capacity. The basic battery
parameters include the nominal voltage $U_{\text{b}}$ (unit: V),
capacity $C_{\text{b}}$ (unit: mAh) and maximum discharge current
$I_{\text{bMax}}$ (unit: A).

(3) \textit{Airframe Diameter}. The airframe diameter $D_{\text{air}}$
(unit: m) is defined as the diameter of the circle formed by the motors,
which is necessary to design an airframe or select a suitable product
on the market. For example, if the obtained airframe diameter is $D_{\text{air}}=0.45\text{m}$,
then the popular airframe \textsl{DJI F450} can be selected to assemble
the multicopter.

(4) \textit{Actual Performance}. The actual performance of the designed
multicopter is necessary for users to evaluate the optimization effect.
In this paper, the actual hovering time, payload weight and thrust
ratio are represented by $\overline{t}_{\text{fly}},\overline{m}_{\text{load}},$
and $\overline{\gamma}$, respectively. The weight of the design multicopter
is represented by $m_{\text{copter}}$ (unit: kg).

For simplicity, the above design outputs are marked with an output
parameter set $\mathbf{\Theta}_{\text{out}}$ as
\begin{equation}
\mathbf{\Theta}_{\text{out}}\triangleq\left\{ \Theta_{\text{m}},\Theta_{\text{e}},\Theta_{\text{p}},\overline{t}_{\text{fly}},\overline{m}_{\text{load}},\overline{\gamma},U_{\text{b}},C_{\text{b}},I_{\text{bMax}},D_{\text{air}},m_{\text{copter}}\right\} .\label{eq:PhiOut}
\end{equation}

\subsection{Optimization Constraints}

The following constraints should be considered to ensure the safety
and practicability of the obtained multicopter design.

(1) \textit{Requirement Constraint}. The actual performance of the
designed multicopter $\overline{t}_{\text{fly}},\overline{m}_{\text{load}},\overline{\gamma}\in\mathbf{\Theta}_{\text{out}}$
should be equal (or close) to the desired performance $\hat{t}_{\text{fly}},\hat{m}_{\text{load}},\hat{\gamma}\in\mathbf{\Theta}_{\text{in}}$.

(2) \textit{Safety Constraint}. The voltages and currents of electronic
components (motors, ESCs, and batteries) should work within the allowed
range to avoid being burnt out. 

(3) \textsl{Compatibility Constraint}. Compatibility is important
during multicopter design. Components have to be compatible with each
other or they cannot work properly or even fail in some cases. Incompatibilities
often occur between ESCs and motors. Consequently, every motor product
lists its compatible ESC products on its description page. 

(4) \textsl{Structure Constraint}. The structure constraint determines
the airframe design. If the airframe is too small, then the propellers
may collide with each other. 

\subsection{Optimization Objectives}

In practice, there is no uniform standard to evaluate the design of
a multicopter. According to \citep{arellano2016multirotor,tian2016mechatronic},
the multicopter design optimization problem is essentially a multiobjective
optimization problem, so an \textquotedblleft optimal design\textquotedblright{}
should fully consider all factors concerned by designers and customers.
Therefore, an objective function $J=f_{J}\left(\mathbf{\Theta}_{\text{out}}\right)$
is proposed in this paper to evaluate the obtained multicopter designs
and find the optimal $\mathbf{\Theta}_{\text{out}}^{*}$.

Compared to the objective functions in \citep{arellano2016multirotor,tian2016mechatronic},
two improvements are introduced in this paper. First, the normalization
operation is applied to each evaluation index, which makes it easier
to determine the weight factor according to the preferences of designers.
Second, more evaluation indexes are considered in this paper according
to actual applications. These evaluation indexes are as follows: 

(1) \textit{Size and Weight}. The size and weight are the most important
factors for designers. A smaller and lighter multicopter is more portable
and convenient for users. 

(2) \textit{Requirement Agreement}. The performance (hovering time,
payload capability, and maneuverability) of the obtained multicopter
should be as close as possible to the given design requirements. 

(3) \textsl{Efficiency}. The efficiency should be as high as possible
to consume less power under the same conditions. 

(4) \textsl{Practicability}. For most users, the components of the
multicopter design should be easy to find on the market. 

(5) \textsl{Safety Margin}. The safety margin indicates that the full-throttle
current of a propulsion system should maintain a certain distance
based on its safety limit, which ensures that the multicopter works
safely in a wider range of flight conditions. 

\subsection{Solving Method}

\begin{figure}[tbh]
\centering \includegraphics[width=0.6\textwidth]{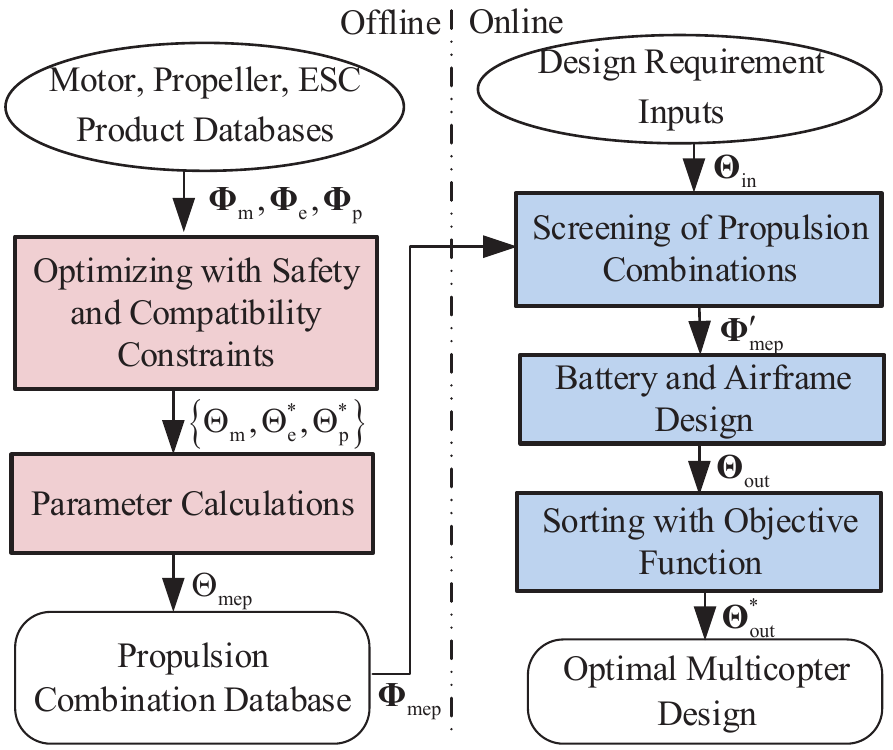}\caption{Framework of the optimization design method.}
\label{Fig03} 
\end{figure}

In summary, the problem inputs are the product databases $\mathbf{\Phi}_{\text{m}},\mathbf{\Phi}_{\text{e}},\mathbf{\Phi}_{\text{p}}$
and design requirements $\mathbf{\Theta}_{\text{in}}$. The problem
output is the optimal multicopter design $\mathbf{\Theta}_{\text{out}}^{*}$.
Fig.\,\ref{Fig03} presents the key steps of the solving method in
this paper. The detailed procedures in Fig.\,\ref{Fig03} will be
presented in \textsl{Section \ref{SEC3}} (offline algorithm) and
\textsl{Section \ref{sec:4}} (online algorithm).

\section{Offline Propulsion System Optimization}

\label{SEC3}

\begin{figure}[tbh]
\centering \includegraphics[width=0.65\textwidth]{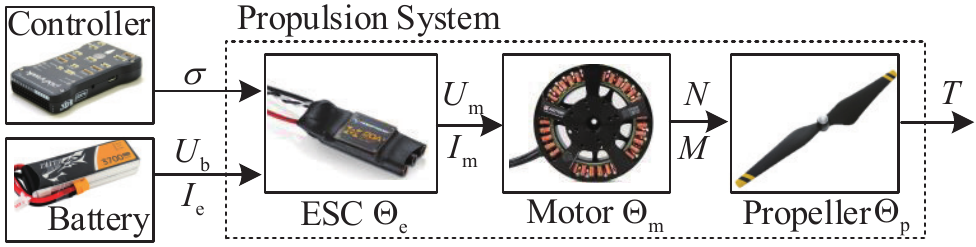}\caption{Connection and power consumption of a propulsion system. }
\label{Fig06} 
\end{figure}

Fig.\,\ref{Fig06} shows the component connection and power consumption
of a propulsion system. The power of a propulsion system is provided
by the battery whose input voltage and current are represented by
$U_{\text{b}}$ (unit: V) and $I_{\text{e}}$ (unit: A) respectively.
The flight controller sends the throttle control signal $\sigma\in\left[0,1\right]$
to ESC, and then ESC generates a PWM-modulated signal to control the
propeller rotating speed $N$ (unit: RPM) to get the desired thrust
$T$ (unit: N). With the throttle signal $\sigma$ increasing from
0 to 1 (full throttle), the values of $T$,$I_{\text{e}}$,$N$ increase
from 0,0,0 to their full-throttle maximum values $T^{*},I_{\text{e}}^{*},N^{*}$.

\subsection{Optimal propulsion Combination}

\subsubsection{Evaluation Indexes}

In practice, the full-throttle thrust $T^{*}$, the thrust efficiency
$\eta^{\text{t}}$ (unit: N/W) and the total weight $m_{\text{mep}}$
(unit: kg) are three most concerned evaluation indexes for propulsion
systems. The thrust efficiency $\eta^{\text{t}}$ is defined as the
ratio between the instantaneous output thrust $T$ and ESC input power
$P_{\text{e}}$
\begin{equation}
\eta^{\text{t}}\triangleq\frac{T}{P_{\text{e}}}=\frac{T}{U_{\text{b}}\cdot I_{\text{e}}}.\label{eq:ThtustEff}
\end{equation}
where $U_{\text{b}}$ (unit: V) is the battery voltage, and $I_{\text{e}}$
is the instantaneous ESC input current. A higher thrust efficiency
$\eta^{\text{t}}$ indicates a lower input power $P_{\text{e}}$ for
generating the same thrust $T$, which leads to longer hovering time.
Noteworthy, the thrust efficiency $\eta^{\text{t}}$ is not a constant
value, and it changes with the throttle $\sigma$. Therefore, the
full-throttle thrust efficiency $\eta^{\text{t*}}=T^{*}/\left(U_{\text{b}}I_{\text{e}}^{*}\right)$
is usually adopted to evaluate the efficiency of a propulsion system. 

\subsubsection{Objective Function}

The objective function to evaluate a motor-ESC-propeller propulsion
combination is given by
\begin{equation}
\begin{array}{cl}
J_{\text{mep}} & =f_{J_{\text{mep}}}\left(\Theta_{\text{m}},\Theta_{\text{e}},\Theta_{\text{p}}\right)\\
 & =k_{\text{m}1}\cdot\frac{T^{*}}{\overline{T}^{*}}+k_{\text{m}2}\cdot\frac{\eta^{\text{t*}}}{\overline{\eta}^{\text{t*}}}+k_{\text{m}3}\cdot\frac{-m_{\text{mep}}}{\overline{m}_{\text{mep}}}
\end{array}\label{eq:Jmep}
\end{equation}
where $\overline{T}^{*},\overline{\eta}^{\text{t*}},\overline{m}_{\text{mep}}$
are normalization parameters, and $k_{\text{m}1},k_{\text{m}2},k_{\text{m}3}$
are weight factors. The normalization parameter $\overline{T}^{*}$
is defined as the maximum full-throttle thrust $T^{*}$ among all
obtained combinations. The same definitions are applied to the normalization
parameters $\overline{\eta}^{\text{t*}},\overline{m}_{\text{mep}}.$
The parameters $k_{\text{m}1},k_{\text{m}2},k_{\text{m}3}$ are positive
coefficients specified by designers according to design preferences.
Usually, $\left\{ k_{\text{m}1},k_{\text{m}2},k_{\text{m}3}\right\} =\left\{ 1,1,1\right\} $
can be selected if there is no special requirement. A better propulsion
combination should have larger $J_{\text{mep}}$ which indicates larger
output thrust $T^{*}$, higher efficiency $\eta^{\text{t*}}$ and
lighter weight $m_{\text{mep}}$. Noteworthy, more evaluation indexes
(cost, size, etc.) can also be included in $J_{\text{mep}}$ in Eq.\,(\ref{eq:Jmep})
according to the actual demand.

\subsubsection{Constraints}

The safety constraint for a propulsion system denotes that the voltage
and current of motors and ESCs should not exceed their allowable limits
for safe operation, which can be described as
\begin{equation}
\begin{array}{c}
I_{\text{e}}^{*}\leq I_{\text{mMax}},I_{\text{e}}^{*}\leq I_{\text{eMax}}\\
U_{\text{b}}\leq U_{\text{mMax}},U_{\text{b}}\leq U_{\text{eMax}}
\end{array}\label{eq:PropConstr}
\end{equation}
where $I_{\text{mMax}},U_{\text{mMax}}\in\Theta_{\text{m}}$ and $I_{\text{eMax}},U_{\text{eMax}}\in\Theta_{\text{e}}$
denote the current and voltage upper limits of the motor and the ESC
respectively.

The compatibility constraint requires that the propulsion components
(motor, ESC and propeller) should be compatible with each other. Since
manufacturers will list the compatible products on product websites,
using a lookup table to store these compatible relationships among
components is a possible way to describe the compatibility constraint.
For example, a compatibility judging function $f_{\text{tab}}\left(\Theta_{\text{m}},\Theta_{\text{e}},\Theta_{\text{p}}\right)$
can be obtained based on the lookup table to describe the compatible
constraint as
\begin{equation}
f_{\text{tab}}\left(\Theta_{\text{m}},\Theta_{\text{e}},\Theta_{\text{p}}\right)=\begin{cases}
\text{false}, & \text{incompatible}\\
\text{true}, & \text{compatible}
\end{cases}.\label{eq:compat}
\end{equation}

\textit{Remark 1}. Considering that the products from the same manufacturer
will be compatible with each other, a more convenient way to avoid
the compatibility problem is to select the motor, ESC and propeller
products from the same manufacturer.

\subsubsection{Brute Force Searching Method}

The brute force searching algorithm is described as follows.

\noindent \rule{1\textwidth}{0.75pt}

\textbf{Algorithm 1} Brute force searching algorithm for optimal propulsion
combination

\noindent \rule{1\textwidth}{0.5pt}

\textbf{Step 1}: For each motor $\Theta_{\text{m}}$ in the motor
database $\mathbf{\Phi}_{\text{m}}$, traverse all ESC and propeller
products in the databases $\mathbf{\Phi}_{\text{e}},\mathbf{\Phi}_{\text{p}}$
to screen out the motor-ESC-propeller combinations $\left\{ \Theta_{\text{m}},\Theta_{\text{e}},\Theta_{\text{p}}\right\} $
that satisfy the constraints in Eqs.\,(\ref{eq:PropConstr})(\ref{eq:compat}).

\textbf{Step 2}: Obtain the values of $U_{\text{b}},I_{\text{e}}^{*},T^{*},m_{\text{mep}},\eta^{\text{t*}}$
for each combination \{$\Theta_{\text{m}},\Theta_{\text{e}},\Theta_{\text{p}}$\}
through experimental measurement or theoretical estimation in \citep{Shi2017}.

\textbf{Step 3}. Obtain the normalization parameters $\overline{T}^{*},\overline{\eta}^{\text{t*}},\overline{m}_{\text{mep}}$
by finding the maximum values of $T^{*},\eta^{\text{t*}},m_{\text{mep}}$
from results in Step 2.

\textbf{Step 4}: Calculate the objective function $J_{\text{mep}}$
for each combination, and select the combination with the maximum
$J_{\text{mep}}$ as the optimal combination ${\Theta_{\text{m}},\Theta_{\text{e}}^{*},\Theta_{\text{p}}^{*}}$,
where ${\Theta_{\text{e}}^{*},\Theta_{\text{p}}^{*}}=\text{argmax}_{\Theta_{\text{e}},\Theta_{\text{p}}}f_{J_{\text{mep}}}\left(\Theta_{\text{m}},\Theta_{\text{e}},\Theta_{\text{p}}\right).$

\textbf{Step 5}: Repeat the above procedures for all motors $\Theta_{\text{m}}$
in database $\mathbf{\Phi}_{\text{m}}$, and a series of propulsion
combinations ${\Theta_{\text{m}},\Theta_{\text{e}}^{*},\Theta_{\text{p}}^{*}}_{k}$
can be obtained.

\noindent \rule{1\textwidth}{0.5pt}

\subsubsection{Analytical Solution Method}

In our previous work \citep{dai2018apractical}, an analytical method
is proposed to find an optimal propulsion combination \{$\Theta_{\text{m}}$,$\Theta_{\text{e}}^{*}$,$\Theta_{\text{p}}^{*}$\}
according to the given full-throttle thrust requirement $T^{*}$.
Therefore, by giving a series of $T^{*}$ with values varying from
0.1N to 100N, the optimization method in \citep{dai2018apractical}
can output a series of propulsion combinations \{$\Theta_{\text{m}}$,$\Theta_{\text{e}}^{*}$,$\Theta_{\text{p}}^{*}$\}$_{k}$
that are capable of covering most common multicopters with normal
size and weight.

\subsubsection{Experimental Selection Method}

\begin{figure}[tbh]
\centering \includegraphics[width=0.65\textwidth]{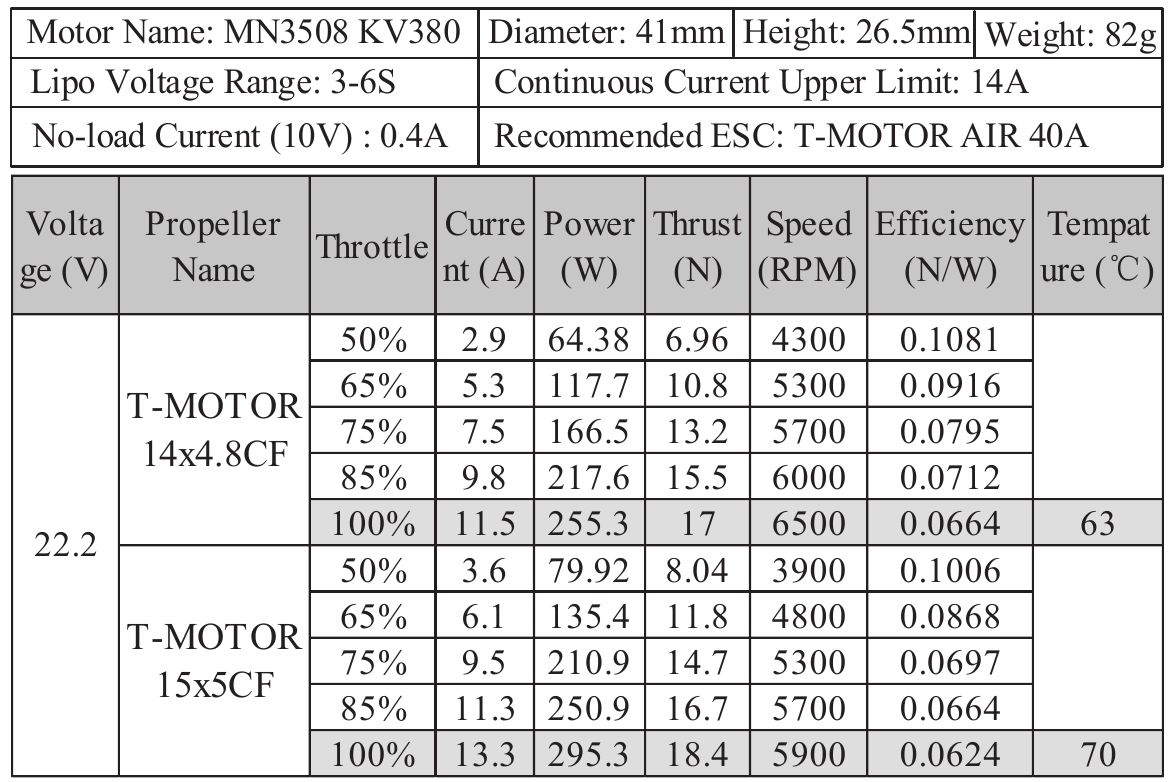}\caption{Product specification and experimental test data of \textsl{T-MOTOR
MN3508 KV380} \citep{TMotor2017}.}
\label{Fig04} 
\end{figure}

Since most manufacturers will publish the motor test data with the
recommended ESC and propeller products, selecting the optimal combination
through these test data is a more convenient and precise way to obtain
the combination database $\mathbf{\Phi}_{\text{mep}}$. For example,
Fig.\,\ref{Fig04} demonstrates the product parameters and test results
of motor MN3508 KV380 on the T-MOTOR website \citep{TMotor2017},
where the recommend two propeller products are 14$\times$4.8CF (Diameter:
14 inches, Pitch: 4.8 inches, Weight: 19.2g) and 15$\times$5CF (Diameter:
15 inches, Pitch: 5 inches, Weight: 26.5g), and the recommend ESC
product is AIR 40 (Maximum Voltage: 22.2V, Maximum Current: 40A, Weight:
26g). Therefore, there are two combinations that can be obtained from
Fig.\,\ref{Fig04}, and the latter one is selected as the optimal
combination according to Eq.\,(\ref{eq:Jmep}), namely \{$\Theta_{\text{m}}$,$\Theta_{\text{e}}^{*}$,$\Theta_{\text{p}}^{*}$\}=\{MN3508
KV380, 15$\times$5CF, AIR 40\}. Since the combination has been tested
and verified through experiments, the safety and compatibility constraints
are strictly satisfied, and the combination is practical enough to
be applied to assemble a real multicopter.

\subsection{Parameter Calculation}

After getting the optimal ESC and propeller for a certain motor, some
parameters are required for the subsequent online design optimization
algorithm including Battery Voltage: $U_{\text{b}}$ (unit: V); Propeller
Diameter: $D_{\text{p}}$ (unit: m); Motor KV Value $K_{\text{V}}$
(unit: RPM/V); Total Weight: $m_{\text{mep}}$ (unit: kg); Full-throttle
Thrust: $T^{*}$ (unit: N); Full-throttle Speed: $N^{*}$ (unit: RPM);
Full-throttle Current: $I_{\text{e}}^{*}$ (unit: A); Air Density:
$\rho$ (unit: kg/$\text{m}^{3}$); Motor Nominal Maximum Current
$I_{\text{mMax}}$ (unit: A). The above parameters can be directly
obtained from the experimental test data as shown in Fig.\,\ref{Fig04}
or calculated with the estimation method in our previous work \citep{Shi2017}.
The air density $\rho$ can be estimated by Eq.\,(\ref{eq:rho})
according to the altitude $h_{\text{t}}$ of the experimental location. 

\begin{figure}[tbh]
\centering \includegraphics[width=0.6\textwidth]{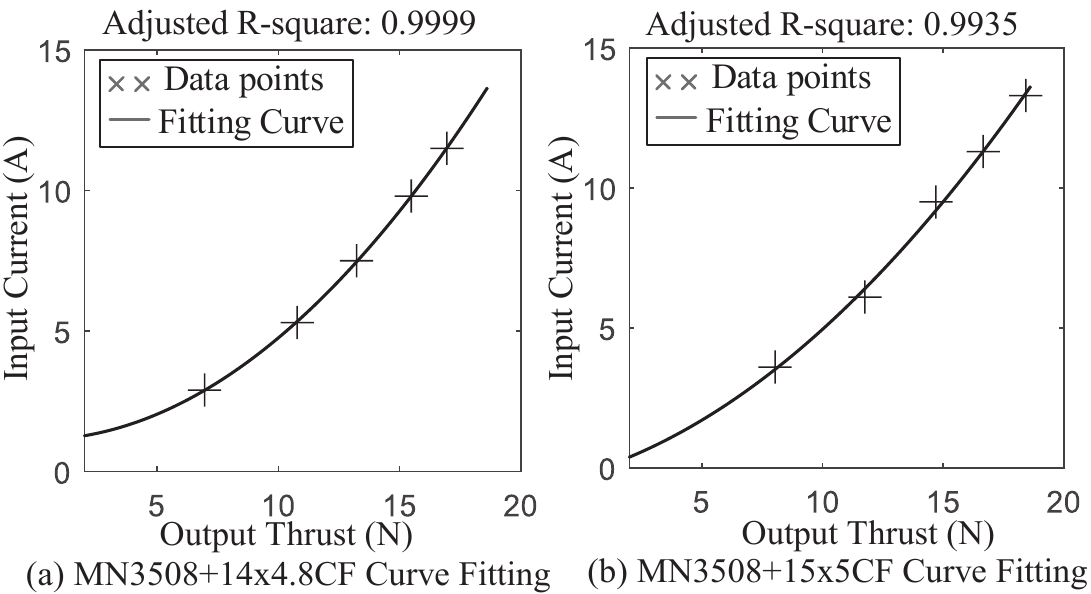}\caption{Second-order polynomial curve fitting results for current-thrust data
of motor \textsl{MN3508}.}
\label{Fig05} 
\end{figure}

According to \citep{Shi2017}, the relationship between the output
thrust $T$ (unit: N) and the input current $I_{\text{e}}$ (unit:
A) (thrust-current curve) of a propulsion combination is very important
for estimating the performance of a multicopter, but it is very complex
and highly nonlinear which requires very accurate parameters and large
computational cost. In order to simplify the computation, the curve
fitting method is adopted to approximate the thrust-current curve
in this paper. Since the curve fitting is finished offline and the
results can be calibrated through experimental test data, the computation
speed and precision of the online multicopter design optimization
algorithm are significantly improved compared with the previous methods.
According to our statistic analysis, the second-order polynomial fitting
method is highly effective in approximating the thrust-current curve
of propulsion combinations, which is given by
\begin{equation}
I_{\text{e}}=f_{\text{IT}}\left(T\right)=k_{\text{t2}}\cdot T^{2}+k_{\text{t1}}\cdot T+k_{\text{t0}}\label{eq:IT}
\end{equation}
where $k_{\text{t0}},k_{\text{t1}},k_{\text{t2}}$ are constant values
obtained by curve fitting. For example, the curve fitting results
for the test data from Fig.\,\ref{Fig04} are presented in Fig.\,\ref{Fig05}.
It can be observed from the figure that both Adjusted R-square values
are larger than 0.99, which indicates that the second-order polynomial
function in Eq.\,(\ref{eq:IT}) can express the thrust-current curve
of a propulsion system in high precision.

\subsection{Database Generation}

The combination \{$\Theta_{\text{m}},\Theta_{\text{e}}^{*},\Theta_{\text{p}}^{*}$\}
along with the necessary parameters are represented by a parameter
set $\Theta_{\text{mep}}$ as
\[
\begin{array}{ll}
\Theta_{\text{mep}}\triangleq & \left\{ \Theta_{\text{m}},\Theta_{\text{e}}^{*},\Theta_{\text{p}}^{*},U_{\text{b}},D_{\text{p}},K_{\text{V}},m_{\text{mep}},\right.\\
 & \left.T^{*},N^{*},I_{\text{e}}^{*},I_{\text{mMax}},\rho,k_{\text{t2}},k_{\text{t1}},k_{\text{t0}}\right\} 
\end{array}.
\]
For example, the optimal propulsion combination for motor \textit{MN3508
KV380} obtained from Fig.\,\ref{Fig04} is $\Theta_{\text{mep}}=$\{\textit{MN3508
KV380}, \textit{15}$\times$\textit{5CF}, \textit{AIR 40}, 22.2, 0.381,
380, 134.5, 18.4, 5900, 13.3, 14, 1.2, 0.0262, 0.2559, -0.2349\}.
By applying the optimal combination method for each motor, a propulsion
combination database $\mathbf{\Phi}_{\text{mep}}$ can be obtained
with $\Theta_{\text{mep}}\in\mathbf{\Phi}_{\text{mep}}$. The online
algorithm will traverse all the combinations $\Theta_{\text{mep}}$
in $\mathbf{\Phi}_{\text{mep}}$ to find the optimal solution for
the desired multicopter.

\textit{Remark 2}. In practice, it is unnecessary to include all the
products on the market to build the database $\mathbf{\Phi}_{\text{mep}}$
because products with similar specifications have similar performance.
According to our practical experience, it is adequate to design common
multicopters (weight range: 0.1kg $\sim$ 50kg) by using a combination
database $\mathbf{\Phi}_{\text{mep}}$ with less than 1000 items.
Of course, professional users can also include their self-designed
products into $\mathbf{\Phi}_{\text{mep}}$ to improve the current
multicopter design.

\section{Online Multicopter Design Optimization}

\label{sec:4}

In this section, for each combination $\Theta_{\text{mep}}\in\mathbf{\Phi}_{\text{mep}}$,
the battery and airframe will be designed according to the design
requirements $\mathbf{\Theta}_{\text{in}}$ to obtain a multicopter
design $\mathbf{\Theta}_{\text{out}}$. Then, an objective function
is proposed to evaluate these multicopter designs to find the optimal
solution $\mathbf{\Theta}_{\text{out}}^{*}$.

\subsection{Selection of Propulsion Combinations}

\subsubsection{Weight Decomposition}

The total weight of a multicopter $m_{\text{copter}}$ (unit: kg)
is given by
\begin{equation}
m_{\text{copter}}=\hat{m}_{\text{load}}+m_{\text{airframe}}+m_{\text{battery}}+n_{\text{p}}\cdot m_{\text{mep}}\label{eq:mTot}
\end{equation}
where $\hat{m}_{\text{load}}\in\mathbf{\Theta}_{\text{in}}$ is the
desired payload weight, $m_{\text{airframe}}$ (unit: kg) is the airframe
weight (includes flight controller and other instruments), $m_{\text{battery}}$
(unit: kg) is the battery weight, $m_{\text{mep}}\in\Theta_{\text{mep}}$
is the weight of a propulsion combination, $n_{\text{p}}\in\mathbf{\Theta}_{\text{in}}$
is the propeller number. Since an airframe is used to support the
whole multicopter weight $m_{\text{copter}}$, according to the statistical
analysis in \citep{Bershadsky2016a}, its weight $m_{\text{airframe}}$
is usually proportional to $m_{\text{copter}}$ as
\begin{equation}
m_{\text{airframe}}=\alpha_{\text{air}}\cdot m_{\text{copter}}\label{eq:mAirf}
\end{equation}
where the airframe weight ratio $\alpha_{\text{air}}$ may vary from
0.08 to 0.40 for multicopters with different materials and structures.
An average value $\alpha_{\text{air}}=0.19$ is suggested in \citep{Bershadsky2016a}
for common multicopters. Meanwhile, by substituting the full-throttle
thrust $T^{*}\in\Theta_{\text{mep}}$ and the input parameters $\hat{\gamma},n_{\text{p}}\in\mathbf{\Theta}_{\text{in}}$
into Eqs.\,(\ref{eq:gama})(\ref{eq:acc1}), the multicopter weight
$m_{\text{copter}}$ is obtained as
\begin{align}
T_{\text{hover}} & =\hat{\gamma}\cdot T^{*}\label{eq:Thover}\\
m_{\text{copter}} & =\frac{n_{\text{p}}\cdot T_{\text{hover}}}{g}=\frac{n_{\text{p}}\cdot\hat{\gamma}\cdot T^{*}}{g}.\label{eq:mCopter1}
\end{align}
Finally, by combining Eqs.\,(\ref{eq:mTot})(\ref{eq:mAirf})(\ref{eq:mCopter1}),
the battery weight $m_{\text{battery}}$ is obtained as
\begin{equation}
m_{\text{battery}}=\left(1-\alpha_{\text{air}}\right)m_{\text{copter}}-\hat{m}_{\text{load}}-n_{\text{p}}\cdot m_{\text{mep}}.\label{eq:mBatterySov}
\end{equation}

\textit{Remark 3}. The full-throttle thrust $T^{*}$ is determined
by the air density $\rho$. Therefore, if the desired air density
$\hat{\rho}\in\mathbf{\Theta}_{\text{in}}$ is not equal to the default
air density $\rho$ in $\Theta_{\text{mep}}$, conversion has to be
made to obtain the actual full-throttle thrust $\overline{T}^{*}$
in Eqs.\,(\ref{eq:mCopter1})(\ref{eq:mCopter1}). Based on the mathematical
model of propulsion systems in \textsl{Appendix A}, the conversion
method to obtain $\overline{T}^{*}$ is derived in \textsl{Appendix
B}. 

\subsubsection{Battery Hovering Current Calculation}

Assuming that the multicopter is in hovering mode, by substituting
Eq.\,(\ref{eq:Thover}) into Eq.\,(\ref{eq:IT}) with $T=T_{\text{hover}}$,
the hovering input current of a propulsion combination $I_{\text{eHover}}$
(unit: A) is given by
\begin{equation}
\begin{array}{c}
I_{\text{eHover}}=k_{\text{t0}}+k_{\text{t1}}\cdot T_{\text{hover}}+k_{\text{t2}}\cdot T_{\text{hover}}^{2}\end{array}.\label{eq:IeHover12}
\end{equation}
Then, the battery hovering current $I_{\text{bHover}}$ (unit: A)
is obtained as
\begin{equation}
I_{\text{bHover}}=n_{\text{p}}\cdot I_{\text{eHover}}+I_{\text{other}}\label{eq:Ib}
\end{equation}
where $n_{\text{p}}\cdot I_{\text{eHover}}$ denotes the total current
of all motors, and $I_{\text{other}}$ (unit: A) denotes other current
consumption such as the current of the flight controller. According
to \citep{Shi2017}, a statistic value $I_{\text{other}}\approx0.5$A
can be applied for the approximate calculation if $I_{\text{other}}$
is unknown.

\textit{Remark 5}. As mentioned before, if the desired air density
$\hat{\rho}\in\mathbf{\Theta}_{\text{in}}$ is different from the
default air density $\rho$ in $\Theta_{\text{mep}}$, conversion
has to be made to obtain the actual input current $\overline{I}_{\text{eHover}}$
in Eq.\,(\ref{eq:IeHover12}), where the conversion method is presented
in \textsl{Appendix C}. 

\subsubsection{Discharge Time Calculation}

The definition of the battery power density $\rho_{\text{b}}$ (unit:
W$\cdot$h/kg) is given by
\begin{equation}
\rho_{\text{b}}\triangleq\frac{U_{\text{b}}\cdot I_{\text{bHover}}\cdot t_{\text{dis}}/60}{m_{\text{battery}}}\label{eq:RhobDef}
\end{equation}
where $t_{\text{dis}}$ (unit: min) is the total discharge time of
a battery, and $U_{\text{b}}\in\Theta_{\text{mep}}$ is the battery
voltage. In practice, a part of the capacity should be retained to
protect the battery from over-discharging. Therefore, the actual battery
discharge time $\overline{t}_{\text{dis}}$ is given by
\begin{equation}
\overline{t}_{\text{dis}}=\alpha_{\text{b}}t_{\text{dis}}=\alpha_{\text{b}}\frac{60\cdot\rho_{\text{b}}m_{\text{battery}}}{U_{\text{b}}\cdot I_{\text{bHover}}}\label{eq:actDis}
\end{equation}
where $\alpha_{\text{b}}\in\left(0,1\right)$ is the discharge capacity
ratio. According to \citep{Shi2017}, an average value $\alpha_{\text{b}}$$\approx$0.9
can be applied for the approximate calculation if $\alpha_{\text{b}}$
is unknown.

\subsubsection{Selection for Design Requirements}

Ideally, the battery discharge time $\overline{t}_{\text{dis}}$ obtained
from Eq.\,(\ref{eq:actDis}) should be equal to the desired hovering
time $\hat{t}_{\text{fly}}\in\mathbf{\Theta}_{\text{in}}$, which
can ensure that the design requirements described by parameters $\hat{m}_{\text{load}},\hat{\gamma},\hat{t}_{\text{fly}}$
are satisfied simultaneously. However, in practice, it is too strict
to find a combination $\Theta_{\text{mep}}$ from database $\mathbf{\Phi}_{\text{mep}}$
that exactly satisfies all requirements $\overline{t}_{\text{dis}}=\hat{t}_{\text{fly}},\overline{m}_{\text{load}}=\hat{m}_{\text{load}},\overline{\gamma}=\hat{\gamma}$.
Therefore, the following selection criterion is proposed to evaluate
whether the propulsion combination satisfies the design requirements
within the tolerable error as
\begin{equation}
\left|\frac{\overline{t}_{\text{dis}}-\hat{t}_{\text{fly}}}{\hat{t}_{\text{fly}}}\right|\leq\varepsilon_{\text{t}}\label{eq:Tdis}
\end{equation}
where $\varepsilon_{\text{t}}$ is a small positive threshold specified
by designers according to the tolerance for design error. If the selection
criterion in Eq.\,(\ref{eq:Tdis}) is satisfied for a motor-ESC-propeller
combination $\Theta_{\text{mep}}$, the actual performance of the
multicopter designed based on $\Theta_{\text{mep}}$ is given by
\begin{equation}
\overline{t}_{\text{fly}}=\overline{t}_{\text{dis}}\approx\hat{t}_{\text{fly}},\overline{m}_{\text{load}}=\hat{m}_{\text{load}},\overline{\gamma}=\hat{\gamma}.\label{eq:tact}
\end{equation}

In summary, the screening algorithm to find the propulsion combinations
that satisfy the given design requirements are listed as follows.

\noindent \rule{1\textwidth}{0.75pt}

\textbf{Algorithm 2} Screening algorithm for requirement-satisfied
combinations

\noindent \rule{1\textwidth}{0.5pt}

\textbf{Step 1}: For a combination $\Theta_{\text{mep}}$ in database
$\mathbf{\Phi}_{\text{mep}}$, the discharge time $\overline{t}_{\text{dis}}$
can be obtained by substituting parameter sets $\Theta_{\text{mep}}$
and $\mathbf{\Theta}_{\text{in}}$ into Eqs.\,(\ref{eq:mCopter1})-(\ref{eq:actDis}). 

\textbf{Step 2}: If the obtained $\overline{t}_{\text{dis}}$ satisfies
the screening criterion in Eq.\,(\ref{eq:Tdis}), then store this
combination $\Theta_{\text{mep}}$ into a database which is marked
as $\mathbf{\Phi}_{\text{mep}}^{\prime}$. 

\textbf{Step 3}: Repeat the above two steps for all combinations $\Theta_{\text{mep}}$
in $\mathbf{\Phi}_{\text{mep}}$, and a database $\mathbf{\Phi}_{\text{mep}}^{\prime}$
is obtained for the following design and optimization procedures.
If no combination is obtained, then an error should be emitted to
stop the whole optimization program.

\noindent \rule{1\textwidth}{0.5pt}

\textit{Remark 6}. The selection criterion in Eq.\,(\ref{eq:Tdis})
can also be formulated based on the errors of $\hat{m}_{\text{load}}$
or $\hat{\gamma}$ instead of $\hat{t}_{\text{fly}}$, which provides
similar selection effect.

\subsection{Battery and Airframe Design}

\subsubsection{Battery Design}

The basic battery parameters that determine a battery product include
the nominal voltage $U_{\text{b}}$ (unit: V), the capacity $C_{\text{b}}$
(unit: mAh) and the maximum discharge current $I_{\text{bMax}}$ (unit:
A). First, the battery voltage $U_{\text{b}}$ can be obtained from
parameter set $\Theta_{\text{mep}}$. Secondly, according to Eq.\,(\ref{eq:Ib}),
the full-throttle battery discharge current is obtained as $I_{\text{b}}^{*}=n_{\text{p}}\cdot I_{\text{e}}^{*}+I_{\text{other}}$,
where $I_{\text{e}}^{*}\in\Theta_{\text{mep}}$ is the full-throttle
current of a propulsion system. It is required that $I_{\text{bMax}}\geq I_{\text{b}}^{*}$
for the safe operation of batteries. With necessary safety margin,
the battery maximum discharge current $I_{\text{bMax}}$ is obtained
as 
\begin{equation}
I_{\text{bMax}}=\alpha_{\text{Ib}}\cdot I_{\text{b}}^{*}=\alpha_{\text{Ib}}\cdot\left(n_{\text{p}}\cdot I_{\text{e}}^{*}+I_{\text{other}}\right)
\end{equation}
where usually $\alpha_{\text{Ib}}\geq1.5$. Thirdly, according to
the definition of the battery capacity with unit mAh, the battery
$\text{\ensuremath{C_{\text{b}}}}$ is obtained as
\begin{equation}
C_{\text{b}}=1000\cdot I_{\text{bHover}}\frac{\overline{t}_{\text{fly}}/\alpha_{\text{b}}}{60}
\end{equation}
where $I_{\text{bHover}}$ and $\alpha_{\text{b}}$ are presented
in Eq.\,(\ref{eq:Ib})(\ref{eq:actDis}). Thus, the battery parameters
$U_{\text{b}},I_{\text{bMax}},C_{\text{b}}$ are obtained for the
output parameter set $\Theta_{\text{out}}$.

\subsubsection{Airframe Design}

\begin{figure}[tbh]
\centering \includegraphics[width=0.65\textwidth]{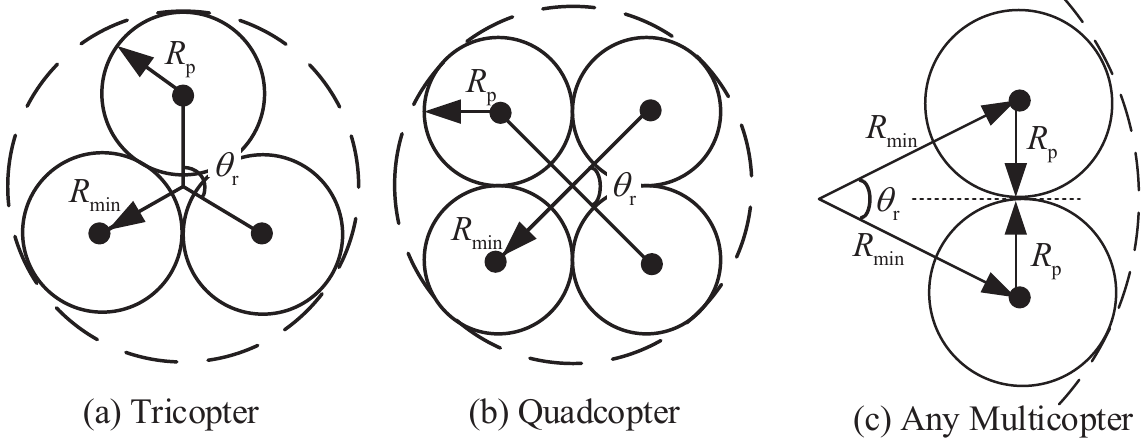}\caption{Minimum airframe for multicopter design.}
\label{Fig07} 
\end{figure}

The airframe diameter $D_{\text{air}}$ (unit: m) is the most important
parameter in designing or selecting an airframe product. In practice,
$D_{\text{air}}$ (unit: m) is expected to be as small as possible.
However, the propellers may interfere with each other if $D_{\text{air}}$
is too small. Fig.\,\ref{Fig07} shows the minimum airframe radius
$R_{\text{min}}$ for different types of multicopters, where $R_{\text{p}}=D_{\text{p}}/2$
is the propeller radius, $R_{\text{air}}=D_{\text{air}}/2$ is the
airframe radius, $\theta_{\text{r}}=2\pi/n_{\text{r}}$ is the angle
(unit: rad) between adjacent arms, and $n_{\text{r}}$ is the number
of arms. Therefore, according to the geometrical relationship in Fig.\,\ref{Fig07},
the minimum airframe radius $R_{\text{min}}$ is given by
\begin{equation}
R_{\text{air}}\geq R_{\text{min}}=\frac{R_{\text{p}}}{\sin\left(\frac{\theta_{\text{r}}}{2}\right)}=\frac{D_{\text{p}}}{2\sin\left(\pi/n_{\text{r}}\right)}.
\end{equation}
Although the constraint $R_{\text{air}}\geq R_{\text{min}}$ can avoid
physical interference between two adjacent propellers, the aerodynamic
interference still exists when two propellers are too close to each
other \citep{Harrington2011}. Therefore, necessary gap is required
between adjacent propellers, so the optimal airframe diameter $D_{\text{air}}$
can be selected as
\begin{equation}
D_{\text{air}}=2R_{\text{air}}=2\alpha_{\text{r}}\cdot R_{\text{min}}=\frac{2\alpha_{\text{r}}D_{\text{p}}}{2\sin\left(\pi/n_{\text{r}}\right)}
\end{equation}
where $\alpha_{\text{r}}=1.05\sim1.2$ is suggested in \citep[p. 62]{quan2017introduction}. 

\subsection{Evaluation, Ordering and Optimization}

Repeating the above procedures for all combinations $\Theta_{\text{mep}}$
in database $\mathbf{\Phi}_{\text{mep}}^{\prime}$, their corresponding
airframe and battery parameters can be obtained according to the requirement
and structure constraints. As a result, a series of multicopter designs
$\Theta_{\text{out}}$ defined in Eq.\,(\ref{eq:PhiOut}) are obtained.
In this subsection, an objective function $J=f_{J}\left(\Theta_{\text{out}}\right)$
is proposed to evaluate these multicopter designs $\Theta_{\text{out}}$.
The one with the minimum $J$ will be selected as the optimal multicopter
design $\Theta_{\text{out}}^{*}$. 

The objective function $f_{J}\left(\Theta_{\text{out}}\right)$ is
given by
\begin{equation}
J=f_{J}\left(\Theta_{\text{out}}\right)=\sum_{i=1}^{7}k_{i}\frac{X_{i}}{\overline{X}_{i}}\label{eq:Cost-1}
\end{equation}
where $X_{1},\cdots,X_{7}\in\mathbb{R^{\mathnormal{+}}}$ are evaluation
indexes, $\overline{X}_{1},\cdots,\overline{X}_{7}\in\mathbb{R^{\mathnormal{+}}}$
are normalizing parameters, and $k_{1},\cdots,k_{7}\in\mathbb{R^{\mathnormal{+}}}$
are weight factors. The detailed definitions are given below
\begin{align}
X_{1} & =D_{\text{air}},X_{2}=m_{\text{copter}},\nonumber \\
X_{3} & =\begin{array}{c}
\sqrt{\left(\frac{\overline{t}_{\text{fly}}-\hat{t}_{\text{fly}}}{\hat{t}_{\text{fly}}}\right)^{2}+\left(\frac{\overline{m}_{\text{load}}-\hat{m}_{\text{load}}}{\hat{m}_{\text{load}}}\right)^{2}+\left(\frac{\overline{\gamma}-\hat{\gamma}}{\hat{\gamma}}\right)^{2}}\end{array}\\
X_{4} & =\frac{U_{\text{b}}\cdot I_{\text{eHover}}}{T_{\text{hover}}},X_{5}=U_{\text{b}},X_{6}=C_{\text{b}},X_{7}=\frac{I_{\text{e}}^{*}}{I_{\text{mMax}}}\nonumber 
\end{align}
where $X_{1},X_{2}$ are indexes for the size and weight of the designed
multicopter; $X_{3}$ denotes the matching degree between the actual
performances $\overline{t}_{\text{fly}},\overline{m}_{\text{load}},\overline{\gamma}$
and the desired performances $\hat{t}_{\text{fly}},\hat{m}_{\text{load}},\hat{\gamma}$;
$X_{4}$ is the inverse value of the thrust efficiency as defined
in Eq.\,(\ref{eq:ThtustEff}) in the hovering mode, where a smaller
$X_{4}$ indicates a higher efficiency; $X_{5},X_{6}$ are the evaluation
indexes for the cost and practicability because the products with
larger battery voltage and capacity are more expensive and harder
to find on the market; $X_{7}$ is the ratio between the full-throttle
current and the motor upper limit current, where a smaller value of
$X_{7}$ indicates larger safety margin to protect the motor from
overheating. In summary, the optimal multicopter design is obtained
as $\Theta_{\text{out}}^{*}=\text{argmin}_{\Theta_{\text{out}}}f_{J}\left(\Theta_{\text{out}}\right).$

\textit{Remark 7}. The normalizing parameters $\overline{X}_{1},\cdots,\overline{X}_{7}$
should be specified by designers based on the standard design of the
design requirements in $\mathbf{\Theta}_{\text{in}}$. For example,
for multicopters with desired weight $\hat{m}_{\text{load}}=0.4\sim0.6$kg
and hovering time $\overline{t}_{\text{fly}}=15\sim25$min, a set
of normalizing parameters $\left\{ \overline{X}_{1},\cdots,\overline{X}_{7}\right\} =\left\{ 0.45,1.5,1,11.5,12,5000,0.65\right\} $
can be selected according to \textsl{DJI PHANTOM} quadcopter. In practice,
it is enough to cover common multicopters from 0.1kg to 50kg with
less than 50 normalizing parameter sets.

\textit{Remark 8.} The default weight factors $k_{1},\cdots,k_{7}$
are all equal to 1, but they can be specified by designers for specific
design preferences. For example, for consumer multicopters, the size,
weight and cost are the most concerned indexes, so $\left\{ k_{1},\cdots,k_{7}\right\} =\left\{ 1,1,0.5,0.3,1,1,0.3\right\} $
can be selected as the weight factors for the objective function in
Eq.\,(\ref{eq:Cost-1}).

\section{Experiments and Verification}

\label{Sec-5}

\subsection{Propulsion Combination Optimization}

\begin{figure}[tbh]
\centering \includegraphics[width=0.65\textwidth]{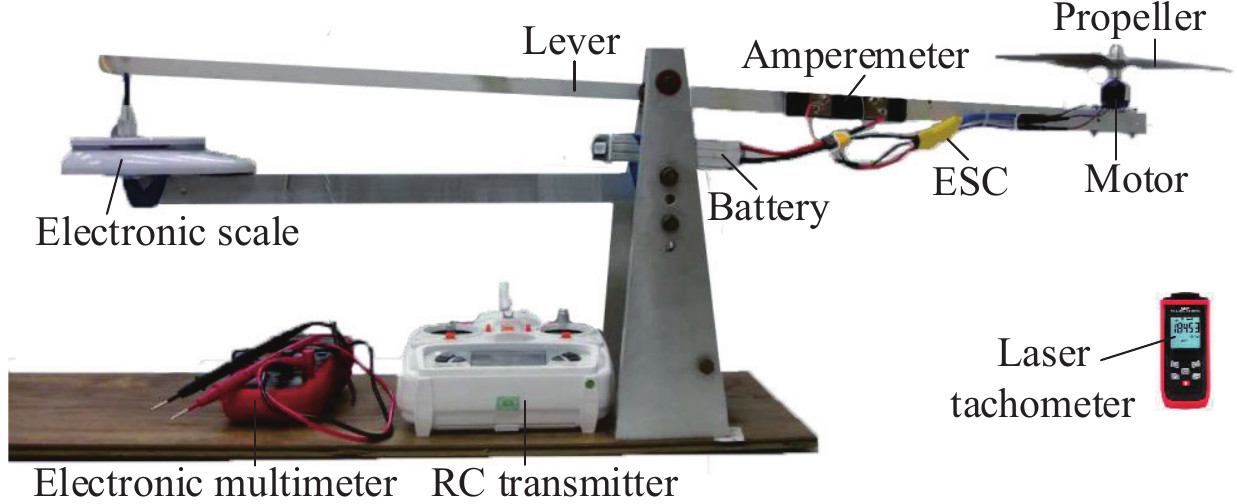}\caption{Indoor measurement device for propulsion systems \citep{Shi2017}.}
\label{Fig08} 
\end{figure}

To measure the performance of a motor-ESC-propeller propulsion combination,
an indoor measurement device (as shown in Fig.\,\ref{Fig08}) was
introduced in our previous work \citep{Shi2017}. The output propeller
thrust $T$, propeller speed $N$, ESC input current $I_{\text{e}}$,
battery voltage $U_{\text{b}}$, and other parameters in Fig.\,\ref{Fig04}
can be easily measured by changing the throttle signal $\sigma$ from
0 to 1. Consequently, by letting output propeller thrust $T$ equal
the hovering thrust $T_{\text{hover}}$ in Eq.\,(\ref{eq:acc1}),
the ESC hovering current $I_{\text{eHover}}$ can be measured through
the device in Fig.\,\ref{Fig08}, and the multicopter hovering time
$\overline{t}_{\text{dis}}$ can be obtained by Eqs.\,(\ref{eq:Ib})
and (\ref{eq:actDis}).

\begin{figure}[tbh]
\centering \includegraphics[width=0.65\textwidth]{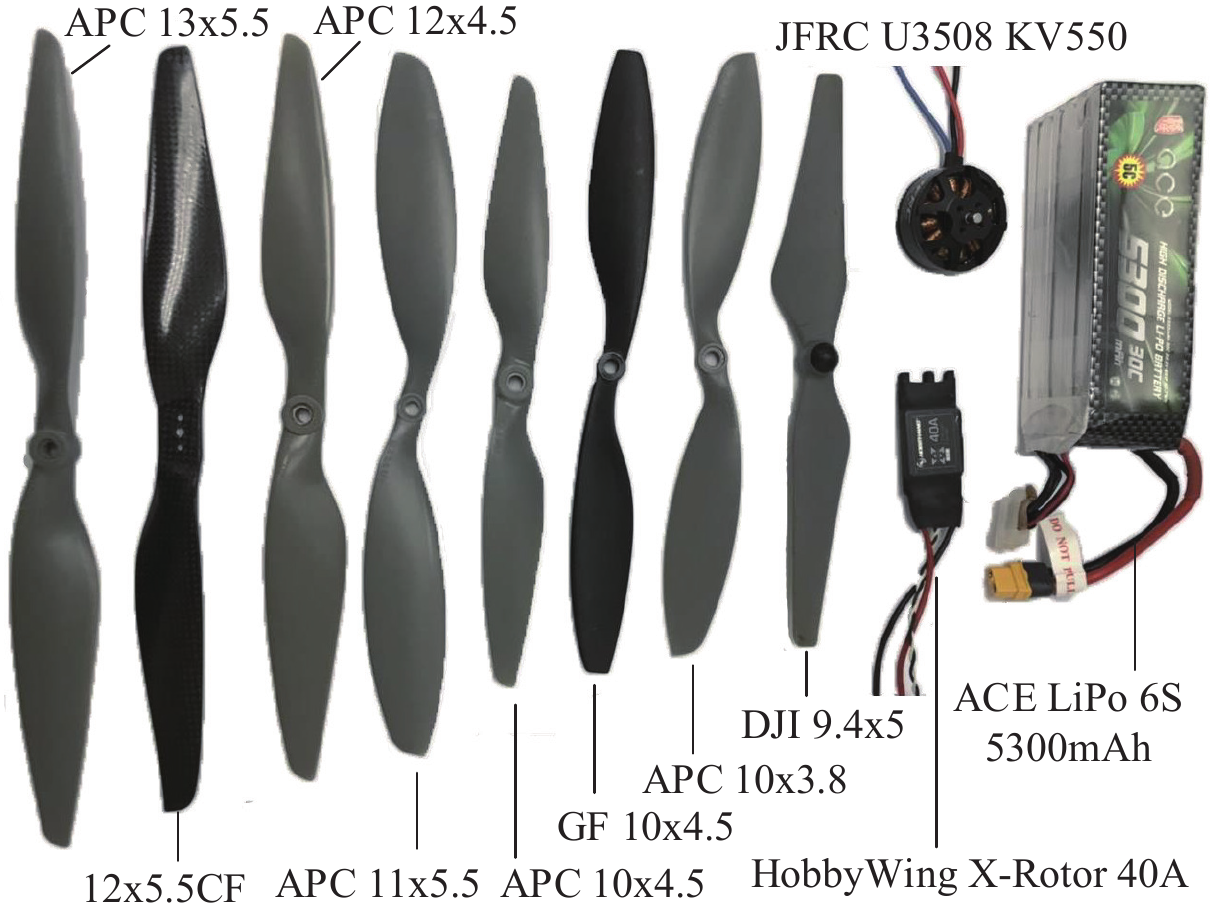}\caption{A test case to find the optimal propeller for a motor JFRC U3508.}
\label{Fig09} 
\end{figure}

A series of tests are performed to find the optimal propeller for
motor JFRC$^{\circledR}$ U3508 KV550. In these tests, as shown in
Fig.\,\ref{Fig09}, the diameters of the selected APC$^{\circledR}$
propellers vary from 9 to 13 in, and the results are listed in Table
\ref{Tab4PropData}. If the propeller is too large (\textgreater{}
12\textit{$\times$}5.5 in Table \ref{Tab4PropData}), then the safety
constraint in Eq.\,(\ref{eq:PropConstr}) is not satisfied because
the full-throttle current $I_{\text{e}}^{*}$ exceeds the motor upper
limit of $20\text{A}$. If the propeller is too small (\textless{}
11$\times$4.5 in Table \ref{Tab4PropData}), the full-throttle thrusts
$T^{*}$ is too low which leads to an exceedingly small $J_{\text{mep}}$.
Therefore, only two propellers (APC 11$\times$5.5 and APC 12$\times$4.5)
are close to the optimal solution. Their detailed measurement results
are listed in the second and third rows of Table \ref{Tab4PropData}. 

According to the optimization steps, the maximum values for $T^{*},\eta^{\text{t*}}$,
and $m_{\text{mep}}$ obtained from Table \ref{Tab4PropData} are
$\overline{T}^{*}=17.84,\overline{\eta}^{\text{t*}}=0.053,$ and $\overline{m}_{\text{mep}}=0.145$.
By specifying the weight factors as $\left\{ k_{\text{m}1},k_{\text{m}2},k_{\text{m}3}\right\} =\left\{ 1,1,1\right\} $,
the values of $J_{\text{mep}}$ are obtained with the objective function
in Eq.\,(\ref{eq:Jmep}). The corresponding results are listed in
the last column of Table \ref{Tab4PropData}. Therefore, the propeller
APC 12$\times$4.5 is the optimal propeller for motor JFRC U3508 KV550
according to our method. The obtained optimal propeller and motor
combination is in accordance with the recommended result from the
JFRC website, which demonstrates the effectiveness of the proposed
optimal selection method for propulsion combinations.

\begin{table}[ptb]
\caption{The results from motor and propeller tests. The propeller name is
formed by Diameter \textit{$\times$} Pitch (unit: inch).}
\label{Tab4PropData}\centering%
\begin{tabular}{|c|>{\centering}p{0.08\textwidth}|>{\centering}p{0.08\textwidth}|>{\centering}p{0.08\textwidth}|>{\centering}p{0.08\textwidth}|>{\centering}p{0.11\textwidth}|>{\centering}p{0.08\textwidth}|}
\hline 
Propellers & $U_{\text{b}}$(V) & $I_{\text{e}}^{*}$(A) & $T^{*}$(N) & $\eta^{\text{t*}}$ & $m_{\text{mep}}$(kg) & $J_{\text{mep}}$\tabularnewline
\hline 
$\leq$ 11$\times$4.5 & - & - & \textless{} 15 & - & - & \textless{} 0.7\tabularnewline
\hline 
\textit{APC} 11$\times$5.5 & 22.2 & 15.5 & 17.84 & 0.053 & 0.141 & 0.85 \tabularnewline
\hline 
\textit{APC} 12$\times$4.5 & 22.2 & 19 & 20.87  & 0.051 & 0.145 & 0.94 \tabularnewline
\hline 
$\geq$ 12$\times$5.5 & - & \textgreater{} 20 & - & - & - & -\tabularnewline
\hline 
\end{tabular}
\end{table}

In addition to using experimental methods to measure the performance
of propulsion combinations, we can use theoretical estimation methods
proposed in \citep{Shi2017,dai2018apractical}. The theoretical methods
are simpler and more convenient, but their calculation precision is
lower than those of the experimental methods, especially when the
product parameters provided by the manufacturers are not sufficiently
precise. By combining the advantages of both methods, designers can
utilize theoretical methods to narrow the search scope, and then use
the experimental methods to find the optimal propulsion combination.
To make it more convenient for readers to apply the proposed method
to multicopter design, a motor-ESC-propeller combination database
was published at the website \uline{\url{http://rfly.buaa.edu.cn/res16/mepdata.xls}}.
This database includes more than 1500 experimentally verified propulsion
combinations, which are adequate for designing multicopters with weights
ranging from 0.2\,kg to 50\,kg.

\subsection{Method Implementation and Comparison}

To make it simpler and more convenient for common users to assemble
their desired multicopters, we developed an online toolbox based on
the proposed multicopter design optimization method. This toolbox
was published online at
\begin{center}
\uline{\url{http://flyeval.com/recalc.html}}
\par\end{center}

\noindent By simply inputting the design requirements, the toolbox
outputs the optimal multicopter design including the size, weight,
payload capability, and propulsion component selection. The program
is fast and can finish within 30 ms by using a web server with a simple
configuration (single-core processor and 1 GB of memory). The number
of visitors to our online tool exceeded 10,000 in 2018, and feedback
indicates that the optimization results from the proposed method are
accurate and practical.

\begin{figure}[tbh]
\centering \includegraphics[width=0.75\textwidth]{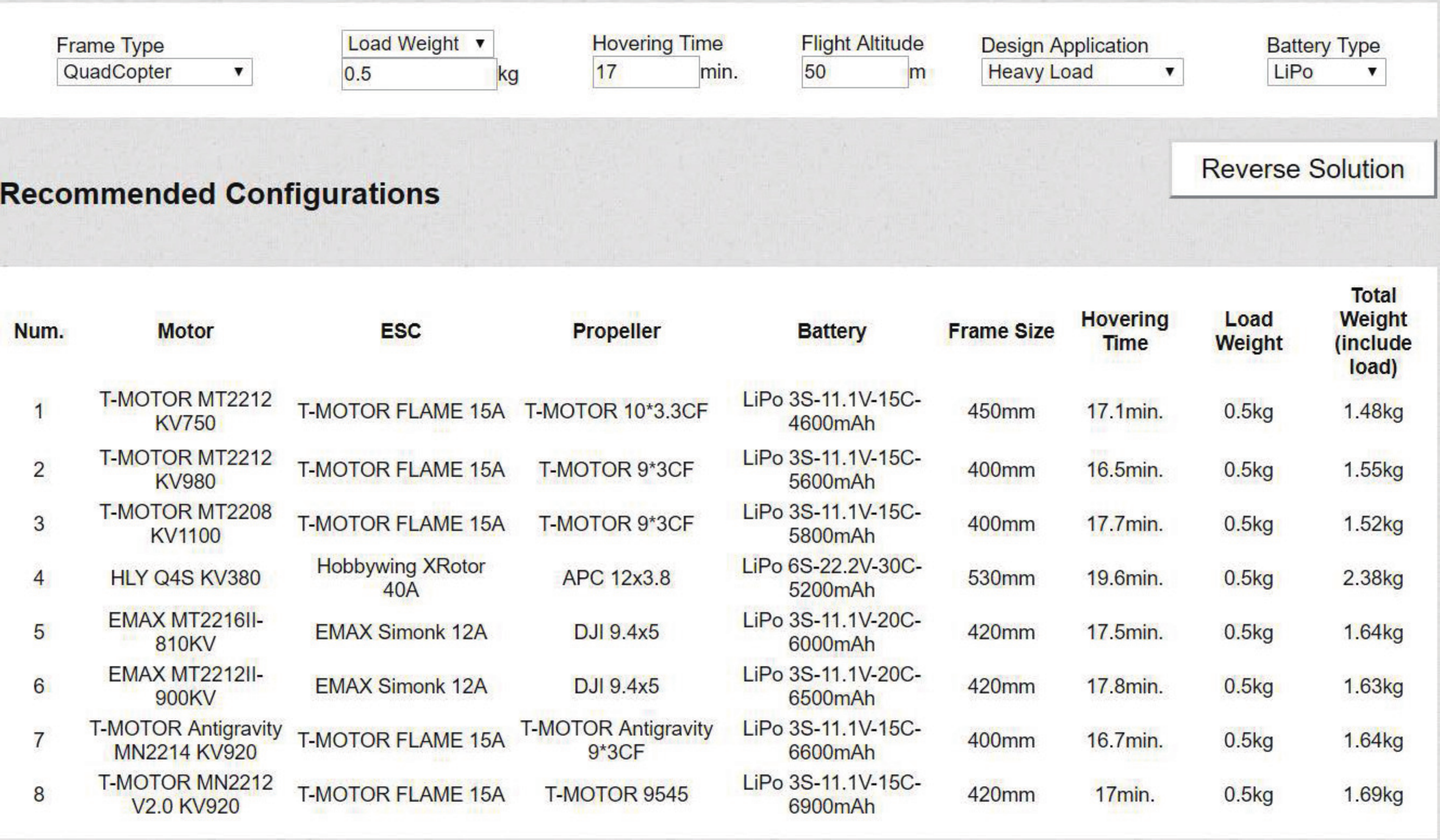}\caption{A calculation case from the online design optimization website.}
\label{Fig10}
\end{figure}

In Fig.\,\ref{Fig10}, an example is presented to design a quadcopter
with the following design requirements: payload weight $\hat{m}_{\text{load}}=0.5$\,kg,
hovering time $\hat{t}_{\text{fly}}=17\text{min}$ and altitude $h=50\text{m}$
(air density $\hat{\rho}=1.22\text{\,kg/m}^{3}$). From our online
toolbox, the thrust ratio is obtained as $\hat{\gamma}=0.55$ according
to the multicopter usage selection of ``heavy load'' (see Fig.\,\ref{Fig10}),
and the battery power density is $\rho_{\text{b}}$$=240$\,W$\cdot$h/kg
with the selection of ``LiPo battery''. The eight most optimal eight
multicopter design results are listed in Fig.\,\ref{Fig10} and are
sorted by the objective function value $J$ in Eq.\,(\ref{eq:Cost-1}).
To verify that the proposed method can obtain multicopter designs
in line with expectations, the input design requirements in Fig.\,\ref{Fig10}
are selected based on a popular quadcopter model \textit{F450} (airframe
diameter: 450mm, weight: 1.5\,kg). 

\begin{figure}[tbh]
\centering \includegraphics[width=0.65\textwidth]{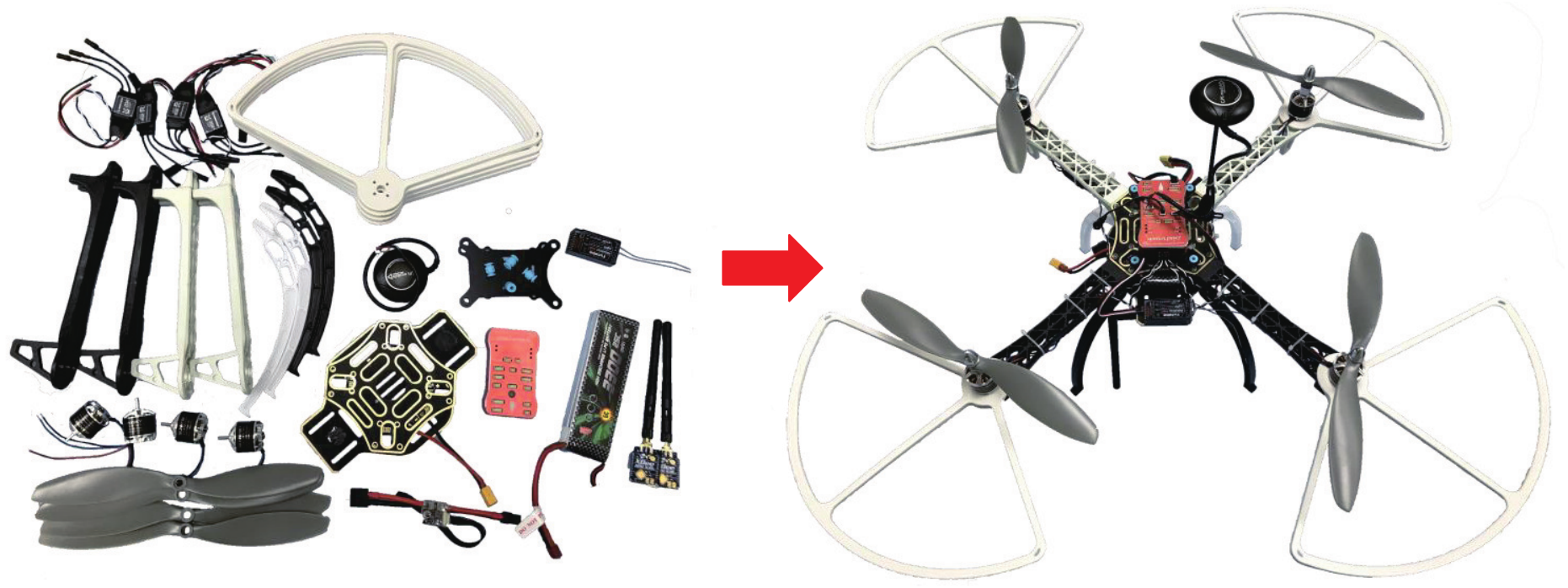}\caption{Assembling a real quadcopter with the obtained design optimization
results.}
\label{Fig11}
\end{figure}

It can be observed from Fig.\,\ref{Fig10} that the diameter and
weight of the obtained multicopter design are approximately 450\,mm
and 1.5\,kg, respectively, which are in line with actual design experience.
For quantitative verification of the experimental results, a test
bench (see Fig.\,\ref{Fig11}) is developed for small-scale multicopters
with different design configurations. The comparison verification
results for the optimal design obtained in Fig.\,\ref{Fig10} are
listed in Table \ref{Tab4Test} along with results from the real multicopter
on the test bench presented in Fig.\,\ref{Fig11}. Since the performance
parameters (e.g., hovering time and payload weight) of a real multicopter
are difficult to measure precisely owing to many actual factors (e.g.,
wind effect, battery states, and controller states), the actual multicopter
performance in Table \ref{Tab4Test} can be considered very close
to the optimization results from the website within the measuring
errors. The comparison results indicate that the obtained multicopter
design is practical for actual multicopter assembly with small estimation
errors.

\begin{table}[ptb]
\caption{Experimental verification for the designed results.}
\label{Tab4Test}\centering%
\begin{tabular}{|c|>{\centering}p{0.1\textwidth}|>{\centering}p{0.1\textwidth}|>{\centering}p{0.1\textwidth}|>{\centering}p{0.1\textwidth}|>{\centering}p{0.12\textwidth}|}
\hline 
Copters & Total weight & Hovering time & Payload Weight & Airframe Size & Battery Capacity\tabularnewline
\hline 
Website  & 1.48\,kg & 17.1\,min. & 0.5\,kg & 450\,mm & 4600\,mAh\tabularnewline
\hline 
Actual  & 1.55\,kg & 18\,min. & 0.55\,kg & 450\,mm & 5000\,mAh\tabularnewline
\hline 
\end{tabular}
\end{table}

In the traditional brutal search method, the computation speed is
usually slow because it must traverse all possible multicopter designs
and evaluate their performance to find the optimal one. In addition,
the performance evaluation will also consume much more time than the
proposed method. For example, it takes about 100\,ms for the evaluation
method in \citep{Shi2017} to calculate the performance of each propulsion
combination, while our online algorithm needs less than 1 ms. In our
simulations, the total computation time of the traditional brutal
search method to obtain an optimal multicopter design from a small-scale
database (five motors, five propellers, and five ESCs) is up to 12.5\,s.
In \citep{Magnussen2015,arellano2016multirotor}, measures were adopted
to reduce the computation amount, but at least five seconds was required
with a high-performance computer.

Compared with the above methods, the number of traversed multicopter
designs and the evaluation time for each design are both significantly
reduced in the proposed method. It takes less than 20\,ms to obtain
a solution from a large-scale database (more than 200 motors) with
a low-performance web server. This would take hours and/or days when
using the previous optimization design methods. 

\section{Conclusions and Future Work}

\label{Sec-6}

This paper proposed a novel and practical method to automatically
calculate an optimal multicopter design according to given design
requirements. The proposed method obtains the optimal products from
a database to form the propulsion system, and subsequently calculates
the optimal parameters for the battery and airframe to satisfy the
design requirements. Specifically, the entire method involves the
use of two algorithms: an offline algorithm and an online algorithm.
Because most of the time-intensive computations are completed in the
offline algorithm, the proposed method can obtain an optimal multicopter
design within 20 ms using a large database (more than 2000 items)
with a low-performance web server. This is much faster than previous
optimization design methods. 

Moreover, practical constraints including safety and compatibility
are fully considered during the optimization process such that the
obtained design can be directly applied to assemble a real multicopter.
Additionally, we proposed a method to map the experimental data from
a specific test condition to any other conditions (different altitudes
or temperatures), which is crucial for multicopter designers to reduce
the testing burden. The experimental results and user feedback for
the online toolbox demonstrate the effectiveness and practicability
of our method. Along with aerodynamic modeling methods for the body
and wing, the method can also be extended to fixed-wing aircraft or
other types of aerial vehicles in the future. 

\section*{Appendices}

\subsection*{Appendix A: Modeling of Multicopter Propulsion System}

As shown in Fig.\,\ref{Fig06}, the motor voltage $U_{\text{m}}$
(unit: V) and current $I_{\text{m}}$ (unit: A) are controlled by
the ESC after receiving the battery voltage $U_{\text{b}}$, the ESC
current $I_{\text{e}}$ and the throttle signal $\sigma$. The ESC
is an energy conversion module which can be described as
\begin{align}
U_{\text{m}} & =\sigma U_{\text{b}}\label{eq:UM}\\
\eta_{\text{e}} & =\frac{I_{\text{m}}U_{\text{m}}}{U_{\text{b}}I_{\text{e}}}\label{eq:EtE}
\end{align}
where $\eta_{\text{e}}\in\Theta_{\text{e}}$ is the energy conversion
efficiency of the ESC, and $\eta_{\text{e}}=0.93\sim0.97$ in most
cases.

According to \citep{Shi2017}, the most commonly used multicopter
motors are brushless direct-current motors. The mathematical model
for a direct-current motor \citep{Shi2017,chapman2005electric} in
the steady state (the rotating speed remains constant) is
\begin{align}
M & =K_{\text{T}}\left(I_{\text{m}}-I_{\text{m0}}\right)\label{eq:MKT}\\
U_{\text{m}} & =I_{\text{m}}R_{\text{m}}+K_{\text{E}}N\label{eq:UmIm}
\end{align}
where $M$ (unit: N$\cdot$m) and $N$ (unit: RPM) are the torque
and rotating speed generated by the motor. Parameters $K_{\text{T}}$
and $K_{\text{E}}$ are motor constants defined as
\begin{equation}
\begin{array}{c}
\frac{\pi}{30}K_{\text{T}}=K_{\text{E}}=\frac{{{U_{\text{{m0}}}}-{I_{\text{{m0}}}}{R_{\text{{m}}}}}}{{{K_{\text{{V}}}}{U_{\text{{m0}}}}}}\end{array}\label{eq:KTKE}
\end{equation}
where $U_{\text{{m0}}},I_{\text{{m0}}},R_{\text{{m}}},K_{\text{{V}}}\in\Theta_{\text{m}}$
denote the no-load voltage, no-load current, resistance, and KV value
which are the basic parameters of a motor. 

For multicopters, the most commonly used propellers are fixed-pitch
propellers whose output thrust $T$ (unit: N) can be modeled as \citep{merchant2006propeller,Merrill2011}
\begin{align}
T & =\begin{array}{c}
C_{\text{T}}\rho\left(\frac{N}{60}\right)^{2}D_{\text{p}}^{4}\end{array}\label{eq:TCT}\\
M & =\begin{array}{c}
C_{\text{M}}\rho\left(\frac{N}{60}\right)^{2}D_{\text{p}}^{5}\end{array}\label{eq:MCM}
\end{align}
where $C_{\text{T}},C_{\text{M}}\in\Theta_{\text{p}}$ are the thrust
coefficient and the torque coefficient of a propeller, $D_{\text{p}}\in\Theta_{\text{p}}$(unit:
m) is the propeller diameter, and $\rho$ (unit: kg/$\text{m}^{3}$)
is the air density. 

\subsection*{Appendix B: Full-throttle Thrust Conversion for Air Density}

By combining Eqs.\,(\ref{eq:UM})(\ref{eq:MKT})(\ref{eq:UmIm})(\ref{eq:MCM})
with $\sigma=1$, the motor model in the full-throttle mode is

\begin{equation}
K_{\text{N}}\rho N^{*2}+K_{\text{E}}N^{*}=U_{\text{b}}-I_{\text{m0}}R_{\text{m}}\label{eq:KNRhp}
\end{equation}
where $K_{\text{N}}$ is an unknown positive constant parameter. Since
both $I_{\text{m0}}$ and $R_{\text{m}}$ are very small values \citep{Shi2017},
it is reasonable to assume that
\begin{equation}
I_{\text{m0}}R_{\text{m}}\approx0.\label{eq:ImoRm0}
\end{equation}
Then, substituting Eqs.\,(\ref{eq:KTKE})(\ref{eq:ImoRm0}) into
Eq.\,(\ref{eq:KNRhp}) gives
\begin{equation}
\begin{array}{c}
K_{\text{N}}\rho N^{*2}+\frac{1}{K_{\text{V}}}N^{*}=U_{\text{b}}\end{array}.\label{eq:KNUb}
\end{equation}
Therefore, the parameter $K_{\text{N}}$ can be obtained according
to Eq.\,(\ref{eq:KNUb}) with $\rho$ and $N^{*}$ as
\begin{equation}
\begin{array}{c}
K_{\text{N}}=\frac{K_{\text{V}}U_{\text{b}}-N^{*}}{\rho N^{*2}K_{\text{V}}}\end{array}.\label{eq:KNSol}
\end{equation}
Let $\overline{N}^{*}$ denote the full-throttle rotating speed under
air density $\hat{\rho}$, and substitute $\rho=\hat{\rho}$ and $N^{*}=\overline{N}^{*}$
into Eq.\,(\ref{eq:KNUb}), the positive solution is given by

\begin{equation}
\begin{array}{c}
\overline{N}^{*}=\frac{-1+\sqrt{4K_{\text{V}}^{2}K_{\text{N}}U_{\text{b}}\hat{\rho}-1}}{2K_{\text{V}}K_{\text{N}}\hat{\rho}}\end{array}.\label{eq:NiStar}
\end{equation}
Meanwhile, according to Eq.\,(\ref{eq:TCT}), the following proportional
expression can be obtained
\begin{equation}
\frac{T^{*}}{\overline{T}^{*}}=\frac{\rho N^{*2}}{\hat{\rho}\overline{N}^{*2}}.\label{eq:TRatio}
\end{equation}
Finally, by substituting parameters $\rho,N^{*},T^{*},K_{\text{V}},U_{\text{b}},\hat{\rho}$
into Eqs.\,(\ref{eq:KNSol})(\ref{eq:NiStar})(\ref{eq:TRatio})
successively, the full-throttle thrust $\overline{T}^{*}$ under air
density $\hat{\rho}$ can be obtained as
\begin{equation}
\overline{T}^{*}=\frac{\hat{\rho}\overline{N}^{*2}}{\rho N^{*2}}T^{*}.\label{eq:T1StarSolv}
\end{equation}

\subsection*{Appendix C: Input Current Conversion for Air Density}

Combining Eqs.\,(\ref{eq:TCT})(\ref{eq:MCM}) with $T=T_{\text{hover}}$
and $M=M_{\text{hover}}$ gives
\begin{equation}
\frac{T_{\text{hover}}}{M_{\text{hover}}}=\frac{C_{\text{T}}}{C_{\text{M}}}\frac{1}{D_{\text{p}}}.\label{eq:TMDiv}
\end{equation}
By substituting Eq.\,(\ref{eq:TMDiv}) into Eq.\,(\ref{eq:MKT}),
the motor hovering current $I_{\text{mHover}}$ is obtained as

\begin{equation}
I_{\text{mHover}}=\frac{M_{\text{hover}}}{K_{\text{T}}}+I_{\text{m0}}=\frac{C_{\text{M}}D_{\text{p}}T_{\text{hover}}}{C_{\text{T}}K_{\text{T}}}+I_{\text{m0}}.\label{eq:ImSov1}
\end{equation}
Meanwhile, according to Eq.\,(\ref{eq:TCT}), the propeller hovering
speed $N_{\text{hover}}$ can be obtained with parameters $N^{*},T^{*},T_{\text{hover}}$
as
\begin{equation}
N_{\text{hover}}=N^{*}\sqrt{\frac{T_{\text{hover}}}{T^{*}}}.\label{eq:NHover}
\end{equation}
Similar to Eq.\,(\ref{eq:KNUb}), the hovering speed $N_{\text{hover}}$
and the hovering motor voltage $U_{\text{mHover}}$ satisfy the following
equation
\begin{equation}
U_{\text{mHover}}=K_{\text{N}}\rho N_{\text{hover}}^{2}+\frac{1}{K_{\text{V}}}N_{\text{hover}}\label{eq:Uhover}
\end{equation}
where $K_{\text{N}}$ is obtained from Eq.\,(\ref{eq:KNSol}). Finally,
the input hovering current $I_{\text{eHover}}$ is given by
\begin{equation}
I_{\text{eHover}}=\frac{I_{\text{mHover}}U_{\text{mHover}}}{U_{\text{b}}\eta_{\text{e}}}.\label{eq:IeHover}
\end{equation}

According to Eq.\,(\ref{eq:TCT}), the actual propeller rotating
speed $\overline{N}_{\text{hover}}$ is determined by parameters $N^{*},T^{*},T_{\text{hover}},\rho,\hat{\rho}$
as
\begin{equation}
\begin{array}{c}
\overline{N}_{\text{hover}}=N^{*}\sqrt{\frac{\rho T_{\text{hover}}}{\hat{\rho}T^{*}}}\end{array}.\label{eq:Nhover1}
\end{equation}
Moreover, according to Eq.\,(\ref{eq:IeHover}), one has
\begin{equation}
\frac{I_{\text{eHover}}}{\overline{I}_{\text{eHover}}}=\frac{I_{\text{mHover}}U_{\text{mHover}}}{\overline{I}_{\text{mHover}}\overline{U}_{\text{mHover}}}\label{eq:IeHoverRatio}
\end{equation}
where $\overline{U}_{\text{mHover}},\overline{I}_{\text{mHover}}$
are the actual motor voltage and current which can be obtained by
Eqs.\,(\ref{eq:ImSov1})(\ref{eq:Uhover}) with $N_{\text{hover}}=\overline{N}_{\text{hover}},\rho=\hat{\rho}$.
Finally, the actual input current $\overline{I}_{\text{eHover}}$
under the desired air density $\hat{\rho}$ can be obtained by substituting
parameters $\rho,N^{*},T^{*},K_{\text{V}},U_{\text{b}},\hat{\rho},T_{\text{hover}},I_{\text{eHover}}$
into Eqs.\,(\ref{eq:KNSol})(\ref{eq:NHover})(\ref{eq:Nhover1})
as
\begin{equation}
\overline{I}_{\text{eHover}}=\frac{K_{\text{N}}\hat{\rho}\overline{N}_{\text{hover}}^{2}+\frac{1}{K_{\text{V}}}\overline{N}_{\text{hover}}}{K_{\text{N}}\rho N_{\text{hover}}^{2}+\frac{1}{K_{\text{V}}}N_{\text{hover}}}I_{\text{eHover}}.\label{eq:IhoverSol}
\end{equation}

\bibliographystyle{IEEEtran}
\bibliography{IEEEabrv}

\begin{thebibliography}{10}
\providecommand{\url}[1]{#1}
\csname url@samestyle\endcsname
\providecommand{\newblock}{\relax}
\providecommand{\bibinfo}[2]{#2}
\providecommand{\BIBentrySTDinterwordspacing}{\spaceskip=0pt\relax}
\providecommand{\BIBentryALTinterwordstretchfactor}{4}
\providecommand{\BIBentryALTinterwordspacing}{\spaceskip=\fontdimen2\font plus
\BIBentryALTinterwordstretchfactor\fontdimen3\font minus
  \fontdimen4\font\relax}
\providecommand{\BIBforeignlanguage}[2]{{%
\expandafter\ifx\csname l@#1\endcsname\relax
\typeout{** WARNING: IEEEtran.bst: No hyphenation pattern has been}%
\typeout{** loaded for the language `#1'. Using the pattern for}%
\typeout{** the default language instead.}%
\else
\language=\csname l@#1\endcsname
\fi
#2}}
\providecommand{\BIBdecl}{\relax}
\BIBdecl

\bibitem{Dai2018}
X.~Dai, Q.~Quan, J.~Ren, and K.~Y. Cai, ``Iterative learning control and
  initial value estimation for probe-drogue autonomous aerial refueling of
  {UAVs},'' \emph{Aerospace Science and Technology}, vol. 82-83, pp. 583--593,
  2018.

\bibitem{quan2017introduction}
Q.~Quan, \emph{Introduction to Multicopter Design and Control}.\hskip 1em plus
  0.5em minus 0.4em\relax Springer, Singapore, 2017.

\bibitem{ke2018design}
Y.~Ke, K.~Wang, and B.~M. Chen, ``Design and implementation of a hybrid {UAV}
  with model-based flight capabilities,'' \emph{IEEE/ASME Transactions on
  Mechatronics}, vol.~23, no.~3, pp. 1114--1125, 2018.

\bibitem{santamaria2012model}
D.~Santamar{\'\i}a, F.~Alarc{\'o}n, A.~Jim{\'e}nez, A.~Viguria, M.~B{\'e}jar,
  and A.~Ollero, ``Model-based design, development and validation for {UAS}
  critical software,'' \emph{Journal of Intelligent \& Robotic Systems},
  vol.~65, no. 1-4, pp. 103--114, 2012.

\bibitem{oktay2016simultaneous}
T.~Oktay, M.~Konar, M.~Onay, M.~Aydin, and M.~A. Mohamed, ``Simultaneous small
  {UAV} and autopilot system design,'' \emph{Aircraft Engineering and Aerospace
  Technology}, vol.~88, no.~6, pp. 818--834, 2016.

\bibitem{Riboldi2018}
C.~E. Riboldi, ``An optimal approach to the preliminary design of small
  hybrid-electric aircraft,'' \emph{Aerospace Science and Technology}, vol.~81,
  pp. 14--31, 2018.

\bibitem{Zhang2019}
S.~Zhang, H.~Li, and A.~A. Abbasi, ``Design methodology using characteristic
  parameters control for low reynolds number airfoils,'' \emph{Aerospace
  Science and Technology}, vol.~86, pp. 143--152, 2019.

\bibitem{Yeo2019}
H.~Yeo, ``Design and aeromechanics investigation of compound helicopters,''
  \emph{Aerospace Science and Technology}, vol.~88, pp. 158--173, 2019.

\bibitem{VolkanPehlivanoglu2019}
Y.~{Volkan Pehlivanoglu}, ``Efficient accelerators for {PSO} in an inverse
  design of multi-element airfoils,'' \emph{Aerospace Science and Technology},
  vol.~91, pp. 110--121, 2019.

\bibitem{Vu2019}
N.~A. Vu, D.~K. Dang, and T.~{Le Dinh}, ``Electric propulsion system sizing
  methodology for an agriculture multicopter,'' \emph{Aerospace Science and
  Technology}, vol.~90, pp. 314--326, 2019.

\bibitem{Introduction2011}
H.~Youngren and M.~Chang, ``Test, analysis and design of propeller propulsion
  systems for {MAVs},'' in \emph{49th AIAA Aerosp. Sci. Meeting New Horizons
  Forum Aerosp. Expo.}\hskip 1em plus 0.5em minus 0.4em\relax AIAA 2011-876,
  Jan. 2011.

\bibitem{Deters2014}
R.~W. Deters, G.~K. Ananda, and M.~S. Selig, ``Reynolds number effects on the
  performance of small-scale propellers,'' in \emph{32nd AIAA Applied
  Aerodynamics Conference}.\hskip 1em plus 0.5em minus 0.4em\relax AIAA
  2007-2151, 2014, pp. 1--43.

\bibitem{burt2008electric}
C.~M. Burt, X.~Piao, F.~Gaudi, B.~Busch, and N.~Taufik, ``Electric motor
  efficiency under variable frequencies and loads,'' \emph{Journal of
  irrigation and drainage engineering}, vol. 134, no.~2, pp. 129--136, 2008.

\bibitem{Shi2017}
D.~Shi, X.~Dai, X.~Zhang, and Q.~Quan, ``A practical performance evaluation
  method for electric multicopters,'' \emph{IEEE/ASME Transactions on
  Mechatronics}, vol.~22, no.~3, pp. 1337--1348, 2017.

\bibitem{dai2018apractical}
X.~Dai, Q.~Quan, J.~Ren, and K.-Y. Cai, ``{An Analytical Design Optimization
  Method for Electric Propulsion Systems of Multicopter {UAVs} with Desired
  Hovering Endurance},'' \emph{IEEE/ASME Transactions on Mechatronics},
  vol.~24, no.~1, pp. 228--239, 2019.

\bibitem{dai2018EFF}
X.~Dai, Q.~Quan, J.~Ren, and C.~Kai-Yuan, ``Efficiency optimization and
  component selection for propulsion systems of electric multicopters,''
  \emph{IEEE Transactions on Industrial Electronics}, vol.~66, no.~10, pp.
  7800--7809, 2019.

\bibitem{Magnussen2014}
{\O}.~Magnussen, G.~Hovland, and M.~Ottestad, ``Multicopter {UAV} design
  optimization,'' in \emph{2014 IEEE/ASME 10th International Conference on
  Mechatronic and Embedded Systems and Applications (MESA)}, sep. 2014, pp.
  1--6.

\bibitem{Magnussen2015}
{\O}.~Magnussen, M.~Ottestad, and G.~Hovland, ``Multicopter design optimization
  and validation,'' \emph{Modeling, Identification and Control}, vol.~36,
  no.~2, pp. 67--79, 2015.

\bibitem{arellano2016multirotor}
V.~Arellano-Quintana, E.~Portilla-Flores, E.~Merchan-Cruz, and P.~Nino-Suarez,
  ``Multirotor design optimization using a genetic algorithm,'' in
  \emph{Unmanned Aircraft Systems (ICUAS), 2016 International Conference
  on}.\hskip 1em plus 0.5em minus 0.4em\relax IEEE, 2016, pp. 1313--1318.

\bibitem{tian2016mechatronic}
F.~Tian and M.~Voskuijl, ``Mechatronic design and optimization using
  knowledge-based engineering applied to an inherently unstable and unmanned
  aerial vehicle,'' \emph{IEEE/ASME Transactions on Mechatronics}, vol.~21,
  no.~1, pp. 542--554, 2016.

\bibitem{orsag2012influence}
M.~Orsag and S.~Bogdan, ``Influence of forward and descent flight on quadrotor
  dynamics,'' in \emph{Recent Advances in Aircraft Technology}.\hskip 1em plus
  0.5em minus 0.4em\relax InTech, 2012.

\bibitem{cavcar2000international}
M.~Cavcar, ``The international standard atmosphere {(ISA)},'' \emph{Anadolu
  University, Turkey}, vol.~30, p.~9, 2000.

\bibitem{TMotor2017}
M.~Wu, ``T-motor official website,'' \url{http://store-en.tmotor.com/},
  accessed September 28, 2018.

\bibitem{Bershadsky2016a}
D.~Bershadsky, S.~Haviland, and E.~N. Johnson, ``Electric multirotor {UAV}
  propulsion system sizing for performance prediction and design
  optimization,'' in \emph{57th AIAA/ASCE/AHS/ASC Structures, Structural
  Dynamics, and Materials Conference}.\hskip 1em plus 0.5em minus 0.4em\relax
  AIAA 2016-0581, Jan. 2015.

\bibitem{Harrington2011}
A.~M. Harrington, ``Optimal propulsion system design for a micro quad rotor,''
  Master's thesis, University of Maryland, College Park, 2011.

\bibitem{chapman2005electric}
S.~Chapman, \emph{Electric machinery fundamentals}.\hskip 1em plus 0.5em minus
  0.4em\relax Tata McGraw-Hill Education, 2005.

\bibitem{merchant2006propeller}
M.~Merchant and L.~S. Miller, ``Propeller performance measurement for low
  {Reynolds} number {UAV} applications,'' in \emph{44th AIAA Aerospace Sciences
  Meeting and Exhibit}.\hskip 1em plus 0.5em minus 0.4em\relax AIAA 2006-1127,
  Jan. 2006.

\bibitem{Merrill2011}
R.~S. Merrill, ``Nonlinear aerodynamic corrections to blade element momentum
  modul with validation experiments,'' Utah State University, Tech. Rep. Paper
  67, 2011.

\end{thebibliography}

\end{document}